# Chiro-Optical Structures with Magnetizable Plasmonic Elements for Modulating the Chiral Transmission of Light

Kaysiyavash Kaykavoosi[1, 2], Nicklas Anttu[1], Javad Ahmadi-Shokouh[2], Hamideh Dashti[2], Mario Zapata-Herrera[3, *], Majid Rashidi Huyeh[4, *]

[1] Department of Physics, Faculty of Science and Engineering, Åbo Akademi University, FI-20500 Turku, Finland

[2] Department of Communications Engineering, Faculty of Electrical & Computer Engineering, University of Sistan and Baluchestan, Zahedan, Iran

[3] Donostia International Physics Center (DIPC), Donostia-San Sebastián 20018, Spain

[4] Department of Physics, Faculty of Basic Sciences, University of Sistan and Baluchestan, Zahedan, Iran

* Corresponding Author: Mario Zapata-Herrera, mario.zapatah@ehu.eus

* Corresponding Author: Majid Rashidi Huyeh, majid.rashidi@phys.usb.ac.ir

**Abstract** — Manipulating external stimuli is crucial for enhancing and controlling the optical response in chiral structures. This study introduces chiral structures composed of single, dimer, or trimer arrays of plasmonic and magneto-plasmonic trapezoidal and triangular nanoantennas. Through modeling, we demonstrate how these systems can (i) function as chiral metasurfaces, (ii) modulate chiral light absorption via an external magnetic field, (iii) magnetically enhance the chiro-optical response, and (iv) improve the chiro-optical and magneto-chiral response through structural engineering. By optimizing the nanoantenna geometry, we achieve enhancements larger than $10^2$ in the chiro-optical signal of a magnetized Ni monomer compared to a non-magnetized system. Additionally, we model bimetallic systems, showing a ~2-fold increase in chiro-optical response for hybrid Au and Ni magneto-plasmonic systems compared to purely plasmonic Au systems, while simultaneously enabling magnetic tuning of the optical response. Our results highlight the potential of structurally engineered magneto-plasmonic nanoantennas for advanced applications in chiro-optical sensing and tunable photonic devices.

Keywords: Surface plasmon resonances, Magneto-plasmonics, Magneto-optics, Chiro-optics, Plasmonic nanoantennas.

## 1. Introduction

Chirality is a geometrical property of an object found in both macrostructures and microstructures [1], and plays an important role in chemistry, physics, and biology [2]. A chiral object and its mirror image are called enantiomorphs or, when referring to molecules, enantiomers, often labeled as “left-handed” or “right-handed” [3] which are impossible to be superimposed onto their mirror image with any combination of translational or rotational operations. The exceptional optical features of chiral materials mainly arise from the difference in their optical response to left circularly polarized- (LCP) and right circularly polarized (RCP) light, known as chiro-optical response [4, 5]. This differential optical response of chiral systems to RCP and LCP light results in a non-identical absorption of RCP and LCP light, called circular dichroism (CD) [6–8].

In metamaterials, chirality of structures enables unique optical properties, such as, negative refractive index [9–12], negative permeability [9, 11–13], negative permittivity [9, 11–14], giant optical activity [9, 15, 16], and very large CD [9, 16–19]. There has been growing interest in actively controlling and switching chirality using external stimuli such as electric fields, temperature, and light [20, 21]. Actively tuning chirality could enable a wide range of enhanced properties in reconfigurable photonic devices with new applications, for example, in nanophotonics-based communication platforms [21, 22], chiral plasmonic sensing [23, 24], and high-resolution displaying devices [20]. This versatility of applications makes such systems of high interest for both fundamental research and practical technological advancements.

Magnetic field-dependent optical responses can be added to the plasmonic elements by incorporating magneto-optically active (MOA) ferromagnetic materials [22, 25, 26]. Briefly, magneto-optics (MO) refers to the effect of an externally applied magnetic field on the propagation of light [27, 28]. In magneto-plasmonics, the effects arising from the interplay between plasmonics and MO phenomena are studied, typically occurring in metallic nanostructures, with the aim to control its MO behavior through plasmon excitations and to influence plasmon characteristics using an external magnetic field [28]. Utilizing magneto-plasmonic compounds, structures with chiral properties can be designed in such a way that the light polarization-dependent optical response can be influenced by an externally applied magnetic field with potential applications in various fields, such as, biosensing, data storage, and optical communication. In biosensing, for instance, the ability to detect small changes in the magnetic field can be used for highly sensitive detection of biomolecules [24]. Chiral magneto-plasmonic structures are of special interest for high-density data storage due to their ability to store information in both the magnetic and plasmonic domains [29]. Furthermore, in optical communication, these structures are possible candidates for the development of new types of optical modulators, isolators, and switches [30–33]. Hence, it is foreseeable that plasmonic structures that simultaneously present chiro-optical and MO properties are promising candidates for the development of magnetic field-tunable chiral plasmonic structures with enhanced chiro-optical responses [22, 26, 34, 35]. The ultimate goal of combining chirality and magneto-plasmonics is to evaluate the possibility of having control over the optical response in magneto-chiral structures and tuning this optical response through the applied magnetic field. Herein, tunability is defined as the magnetic control of the chiro-optical response. Varying the external magnetic field actively switches the chiral response at a target wavelength, and modifying the dimensions of the structures provides a route to engineer and optimize the magnetically-controlled chirality [22, 34, 36, 37]. Nickel (Ni), for example, as a ferromagnetic material, can be magnetized with an external magnetic field, which affects its optical response, leading to a difference in the measured CD under opposite magnetic fields, which is known as magnetic CD (MCD) [26, 38]. Measuring both the CD and MCD for a chiral nanostructure may allow us to discriminate the changes induced by the chiral component of the system from the changes induced by the non-chiral part of the system, which could give rise to advanced strategies of chiro-optical sensing using chiral magneto-plasmonics [26]. Overall, the current research trend aims to deepen our understanding on the fundamental properties of chiral magneto-plasmonic systems and explore their potential applications in various fields.

While numerous studies have explored various geometries and possibilities for active magnetic control of chiro-optical structures [22, 26, 39, 40], there are basic geometrical shapes with high potential for chiral or magneto-chiral properties that remain understudied. For instance, while a single or a pair of disks [22, 40], cubes [41–43], or triangles [44] are inherently achiral on their basic symmetrical form, these shapes can exhibit significant chiral behavior by just engineering their size, shape, and periodicity [22, 40–43]. Also, by an appropriate combination of nanoparticles, the choice of elements, and their respective orientation and integration, it is possible to design bimetallic nanoantennas that exhibit both chiral and magnetic chiro-optical responses, due to the breaking of both the mirror and time-reversal symmetry, respectively. For instance, with the combination of a nanocube and a nanotriangle, which are both achiral, it is possible to form a right-angled nanotrapezoid, which is an intrinsically chiral structure.

Here, we designed and investigated several chiral magneto-plasmonic structures through systematic numerical modelling, comprising trapezoidal and triangular nanoantennas made of gold and nickel. We study the chiro-optical responses of these systems through illumination with RCP- and LCP- light. Then, in the presence of a Ni nanoparticle, the effect of applying an external static magnetic field ($H_{DC}$) [45] on the chiral absorption of CPL in these magneto-chiral structures is evaluated. Our results not only confirm the formation of chiral plasmonic systems, but also show that the optimal combination of magneto-optical and plasmonic elements can enhance the chiro-optical response in bimetallic systems, especially in the trapezoid trimer where we can enhance their CD, on the order of ~2-folds compared to an Au monomer. As for the nanoferromagnetic system, it is shown that the effect of applying an external magnetic field yields in a significant enhancement larger than $10^2$ in the chiro-optical response compared to the non-magnetized system. Overall, this systematic study contributes to further understanding of magnetic chiro-optical properties in chiral magneto-plasmonic systems. Our findings can be used to advance plasmon-based chiral nanophotonics, paving the way for their application in enantioselective optical sensing.

The paper is organized as follows. In section 2 we briefly outline the systems under study and the methodology used. Section 3 presents results and their discussion in section 4. The last section provides summary and conclusions

# 2. Methods

## 2.1. Structures and numerical simulations

In our study, we consider four different types of structures (Fig. 1): a trapezoid monomer (Fig 1a), dimer of nanotrapezoid (Fig. 1b), a trimer of right-angle nanotrapezoids (Fig. 1c), and a system comprising three equilateral nanotriangles (Fig. 1d). To induce MO effects in the dimer and trimer structures, one of the constituent particles is made of Ni (grey coloring in Fig. 1). The gaps between the nanoparticles in the dimer (Fig. 1b) and the trimer (Fig. 1c) of trapezoids are considered to be 5 nm or 10 nm. The triangles (Fig. 1d) are arranged to have 5 nm or 10 nm tip-to-tip gaps, resembling two sets of bow-tie nanoantennas. In our modeling, we adopted rounded edges for all the structures with a curvature radius of 2nm, as is described in detail in section S.3. The nanoantennas are placed in an array with period of 120 nm which is large enough to avoid lattice resonances in the lateral directions, with geometrical dimensions of the nanoantennas shown in Fig. 1.

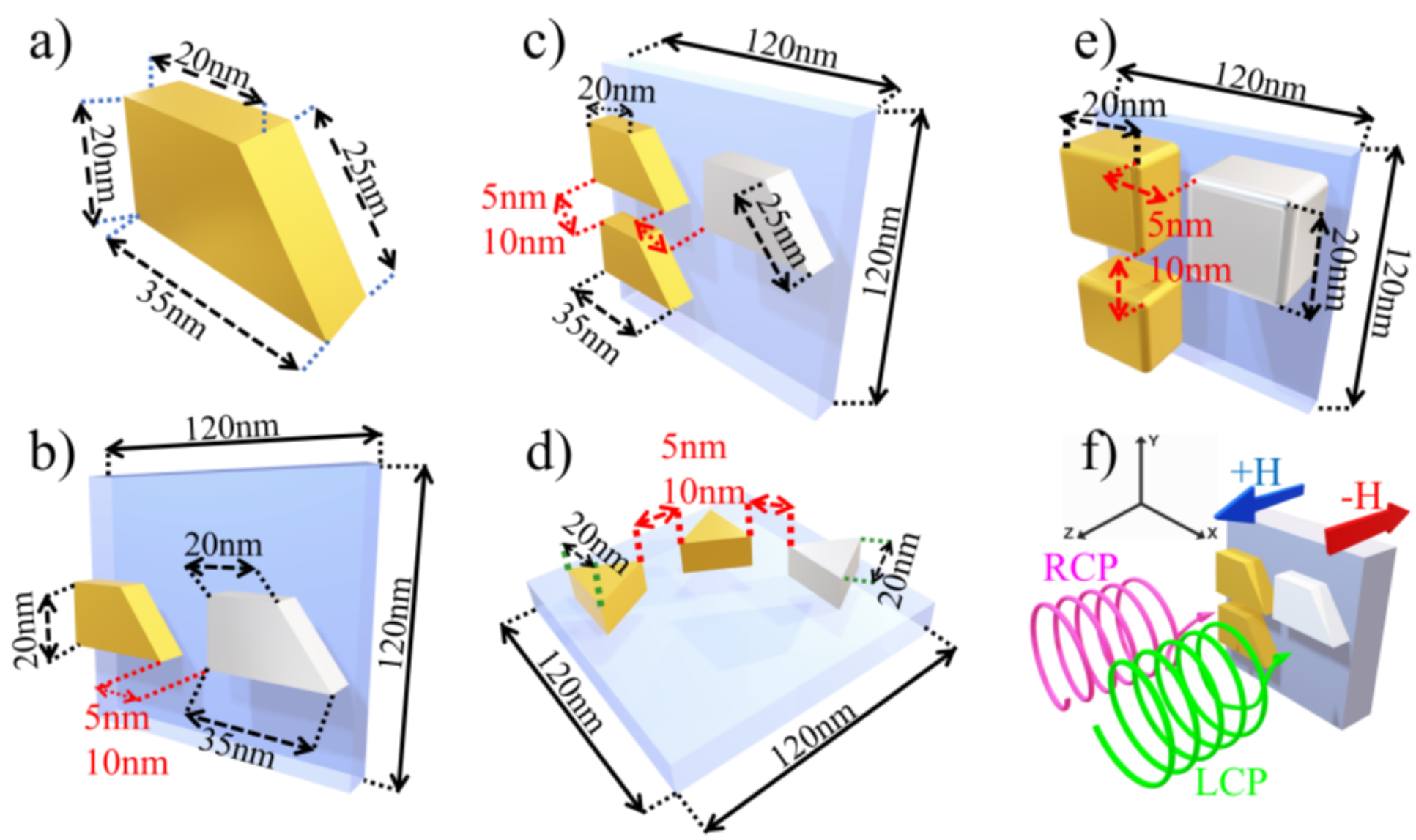

**Fig. 1.** The modeled systems and their respective dimensions; a) monomer, b) dimer, and c) trimer of trapezoids, d) trimer of triangles, and e) trimer of cubes. f) Schematic representation of the CPL illumination and, the magnetically controlled chiral differential response in a trimer system of magneto-plasmonic trapezoidal nanoantennae. (The style of the schematic in Fig .1f is inspired by Refs [22, 46, 47]). The shaded region underneath the nanostructures represents the surface of the substrate in the *xy*-plane.

To model the optical response of the above plasmonic antennas, made of MO and non-MO materials, we used the finite element method (FEM), implemented in the commercial software COMSOL Multiphysics® [48], to obtain $A_{LCP}$ and $A_{RCP}$, the absorbance of LCP and RCP incident light, respectively (see Supporting Information section S.1 and S.2 for details).

The circular dichroism (CD) signal of the systems can be calculated by its differential absorbance $\Delta A = A_{LCP} - A_{RCP}$, as $CD$=$\Delta A$ [4, 5, 49, 50], or by the Kuhn's dissymmetry factor, also known as the *g* factor, directly related to the CD [51]:

$$g = 2\frac{A_{LCP}-A_{RCP}}{A_{LCP}+A_{RCP}}. \quad (1)$$

The *g* factor is especially useful when the chiro-optical response is studied among various kinds of samples, possible varying magnetization conditions and excitation wavelengths, describing the relative polarization-dependent absorption of CPL by a chiral sample [51, 52].

Macroscopic magneto-optical properties are typically taken into account theoretically through the dielectric tensor [53], and the tensor enters the Maxwell's equations that describe the propagation of electromagnetic waves. Hence, in numerical modelling, two magnetic field configurations, namely +*H* and –*H*, were implemented to yield opposite signs for the off-diagonal terms in the Ni permittivity tensor (see Supplementary Section S2 for details). The magnetic fields are assumed to be applied parallel (-*H*) or antiparallel (+*H*) to the light propagation direction (in our case, along the *z*-axis, Fig. 1f) and strong enough to give saturation of the magnetization, *M*, along the normal to the substrate of the system (within the assumption that for $H = 0$, the off-diagonal components are zero) [22, 28, 53–55].

For ferromagnetic systems, the magnetic field induced CD, i.e. the MCD, depends on *M*, which in turn is dependent on *H*, assumed strong enough to saturate *M* as described above, and is defined as [26]:

$$MCD = \Delta CD = CD(+M) - CD(-M). \quad (2)$$

# 3. Results

## 3.1. Monomer of Au and Ni

Since the chiro-optical response is directly derived from the absorbance, it is easily conceivable that the volume of the nanoparticle, among other parameters, has a crucial contribution in determining the chiro-optical properties [56, 57]. The overall optical activity, including circular birefringence and circular dichroism, is affected by the nanoparticle volume. Larger nanoparticles might show different optical features compared to smaller ones due to changes in their electronic and structural properties [58]. However, to function as a metamaterial, the meta-atom, or chiral nanoparticle, should be smaller than the wavelength of the incident light; hence, the size of the chiral nanoparticle is limited to some extent [57].

We calculate the *g* factor for monomer of a Au trapezoid by changing the thickness of the nanoparticle for a systematic choice of an optimized design. The thickness of the Au nanoparticle (*h* in Fig. 2) is varied from 20 nm to 90 nm, in steps of 10 nm.

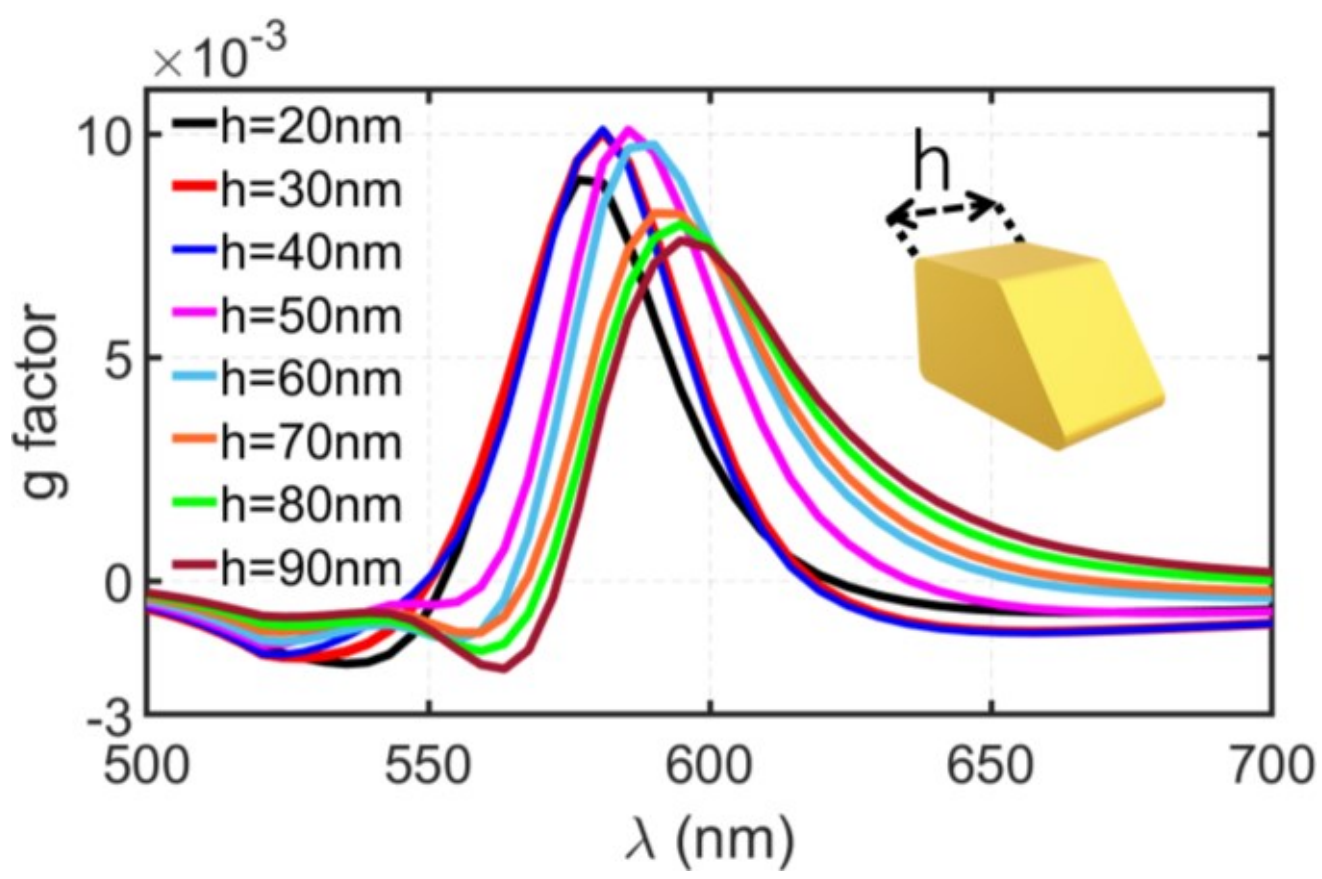

**Fig. 2**. The *g* factor spectra calculated from FEM simulations for a single Au trapezoid with different thicknesses. The corresponding thickness of the studied systems is given as *h*, ranging from 20 nm to 90 nm, with 10 nm steps.

The results in Fig. 2 exhibits an intrinsic chiro-optical activity, but also indicates that as the thickness of the Au nanoparticle is increased, the resonance in the *g* factor spectra exhibits a slight red-shift. Although the magnitude of the *g* factor does not change significantly, the highest values are observed for nanoparticles with thicknesses of 30 nm, 40 nm, and 50 nm. Further increasing of the thickness of this nanoparticle gradually suppresses the chiro-optical response in the studied range. Therefore, we select 50 nm as the thickness for monomers.

In Fig. 3, to compare the effects of choice of the nanoparticle element on the chiro-optical activity in our systems, we calculate CD (Fig. 3a) and the *g* factor (Fig. 3c) for both monomer systems containing either a Au trapezoid, or a magneto-chiral Ni trapezoid. To demonstrate the effects of magnetic field applied in two opposite directions in the Ni nanoantenna, it is shown how the chiro-optical response is undergoing magnetic alternation. We calculate the magnetic tunability of the CD spectra ($MCD = \Delta CD$, Fig. 3b), as well as the magnetic enhancement on the order of ~130-folds in the chiro-optical response as the ratio of the CD response in a magnetized system ($\pm H$) to that of a non-magnetized system (NoH) (Fig. 3d). Thanks to this magnetic enhancement, the Ni trapezoid, under magnetic field, can show comparable chiro-optical responses to that of the Au trapezoid (Fig. 3a,b).

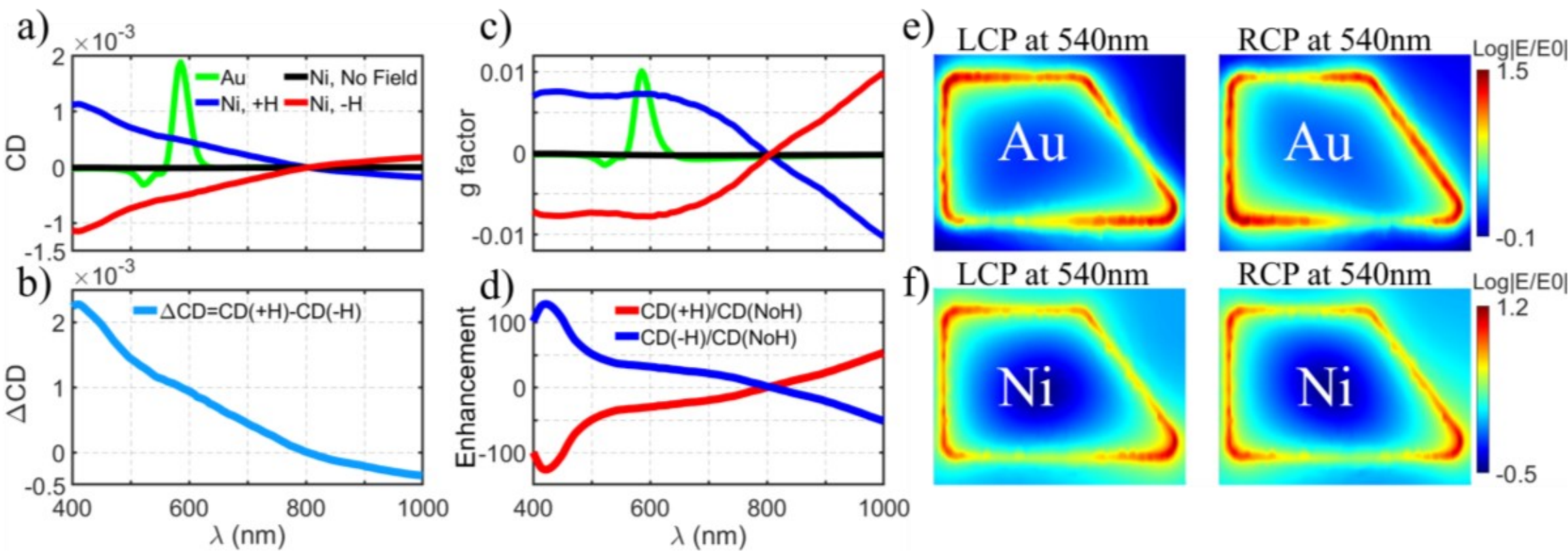

**Fig. 3**. (a) CD, (b) magnetic tunability of CD spectra, as $MCD = \Delta CD$, and (c) $g$ factor, calculated from FEM simulations for a single Au trapezoid (Fig. 3a-c, continuous green lines), and one Ni trapezoid with 50 nm thicknesses, in the absence of an external magnetic field (Fig. 3a-c, No Field, continuous black lines), and in the presence of external magnetic field applied in opposite directions (Fig. 3a-c, $\pm H$, continuous blue and red lines, respectively). and (d) the magnetic enhancement of the chiro-optical response compared to a non-magnetized system of a Ni nanotrapezoid. Normalized electric near-field distribution of a single Au (e) or (f) Ni trapezoid (log scale), upon LCP and RCP light illumination ($\lambda = 540$ nm, chosen based on the far-field transmission spectra in Supporting Information Fig. S3c).

Results presented in Fig. 3 show that this configuration exhibits circular dichroism for Au nanoparticles around the plasmon resonance at $\lambda \approx 550$ nm, showing non-superimposable electric distribution in the *xy*-plane with a top-view when changing the handedness of the CPL. (The electric near-field distributions for monomer structures are given in the Supporting Information Figs. S8–Fig. S12.)

Compared to the CD of a pure Au nanoantenna (Fig. 3a-c), the main effect of using non-magnetized Ni is the significant downscaling of the relative intensity of the chiral response, which can be attributed to the different plasmonic properties of these two elements. This trend could also be seen in the electric field plots of monomers (Fig. 3e-f). Overall, Au nanoparticles are capable of exhibiting stronger chiro-optical response compared to Ni nanoparticles [59–63]. However, the chiro-optical response of the Ni nanoparticle shows to be markedly transformed by the external magnetic field (Fig. 3a-c), promoting the use of such systems for magnetic modulation of chiral response.

## 3.2. Chiro-optical response and MCD in bimetallic magneto-chiral structures

To explore the potential of combining the sharp chiro-optical response of the purely plasmonic Au nanoparticles with the strong magneto-plasmonic response of Ni nanoparticles, we study hybrid systems of Au and Ni chiral magneto-plasmonic nanoparticles. To design an optimal bimetallic system, we analyze the effects of altering structural features on the chiro-optical and magneto-chiroptical responses and behaviors.

As for the choice of an optimized structure, we again study volume effects on the CD and MCD signals of the hybrid Au and Ni chiral magneto-plasmonic nanotrapezoids by changing their thicknesses, and the nanoparticle gaps (Fig. 4).

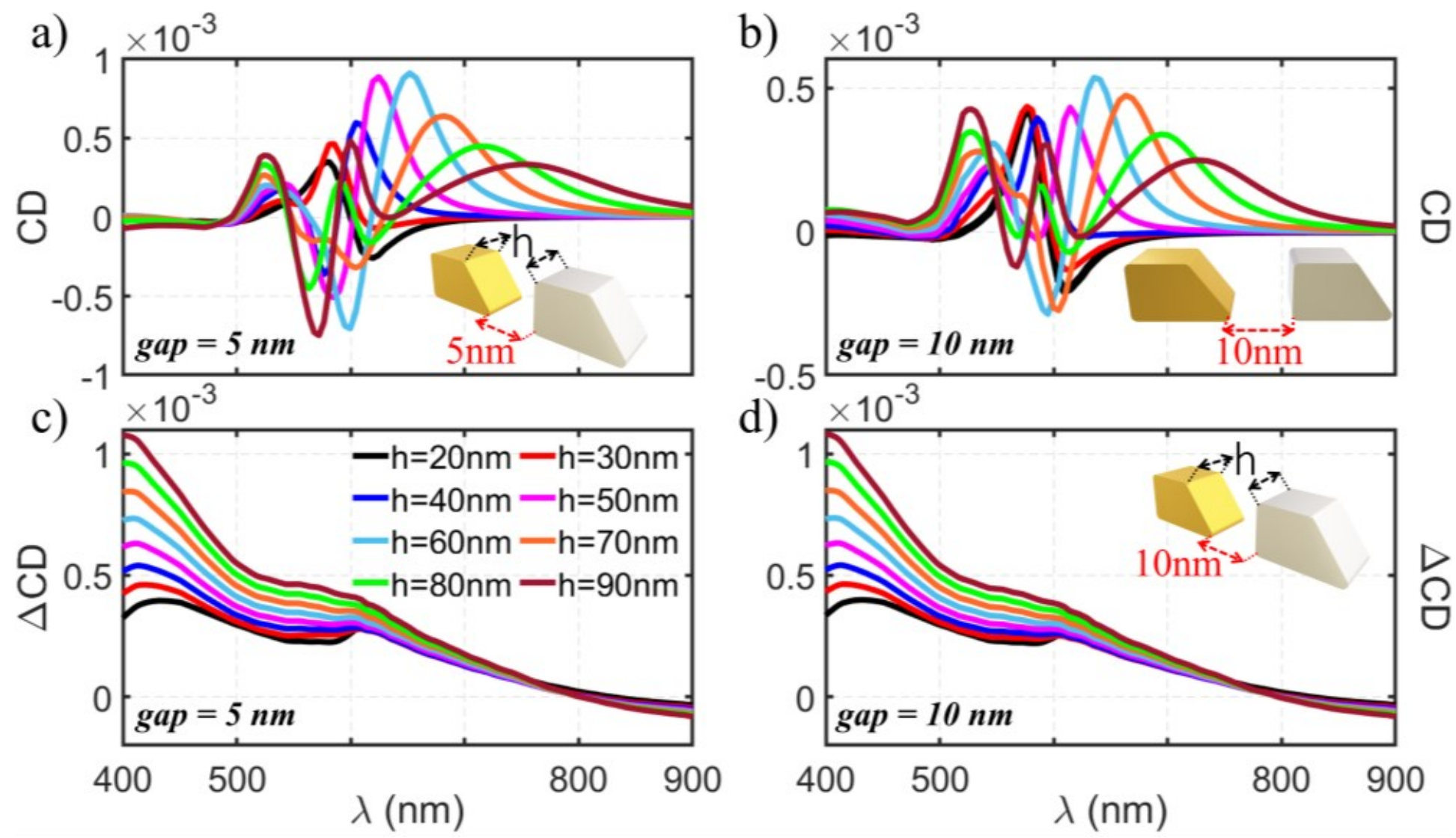

**Fig. 4**. The CD (a.b) and magnetic tunability of the CD spectra (c.d), as $MCD = \Delta CD$, calculated from FEM simulations for a hybrid system of Au and Ni trapezoids with different thicknesses, and different nanoparticle distances. The corresponding thickness of the studied systems is given as $h$, ranging from 20 nm to 90 nm, with 10 nm steps, with gaps of 5 nm (a.c), and 10 nm (b.d).

Remarkably, the optical properties of the chiral nanoparticles (both extinction and chirality) in bimetallic systems also depend strongly on their subwavelength plasmonic gap. It is evident that the magnitude of the CD signal lessens for bigger nanoparticle gaps (Fig. 4a, b). Generally, chiral nanoparticles with narrower and deeper plasmonic gaps exhibit a stronger and redshifted extinction and chiral response [64].

Moreover, in this trapezoidal dimer configuration, the magnitude of the CD signal increases with thickness, peaking at approximately 50 nm in particle height, before declining for thicker structures. Additionally, the resonance broadens and red shifts (Fig. 4a,b). Unlike the CD spectra, which show a strong dependency on the gap between nanoparticles (Fig. 4a,b), the magnetic tunability of the CD response is nearly identical for 5 nm and 10 nm gaps, with its magnitude increasing with thickness (Fig. 4c,d).

Overall, the results in Fig. 4 suggest that designing an optimal chiral magneto-plasmonic system involves a tradeoff between magneto-plasmonic properties. While one system may exhibit the desired CD response, it may not necessarily have strong tunability. Therefore, different applications can determine which enhancements are of interest. We again select a thickness of 50 nm for the nanoparticles in bimetallic hybrid systems, but with gaps of 5 nm and 10 nm.

To further investigate the possibility of designing optimal bimetallic systems with strong hybrid responses and to analyze the effects of altering the structural features on the chiro-optical and magneto-chiroptical properties, in Fig. 5 we plot the *g* factor and ΔCD signals of the proposed hybrid Au and Ni chiral magneto-plasmonic nanoantennas for a comprehensive representation on the choice of the parameters required for different applications (see Fig. 1 for schematic of the systems).

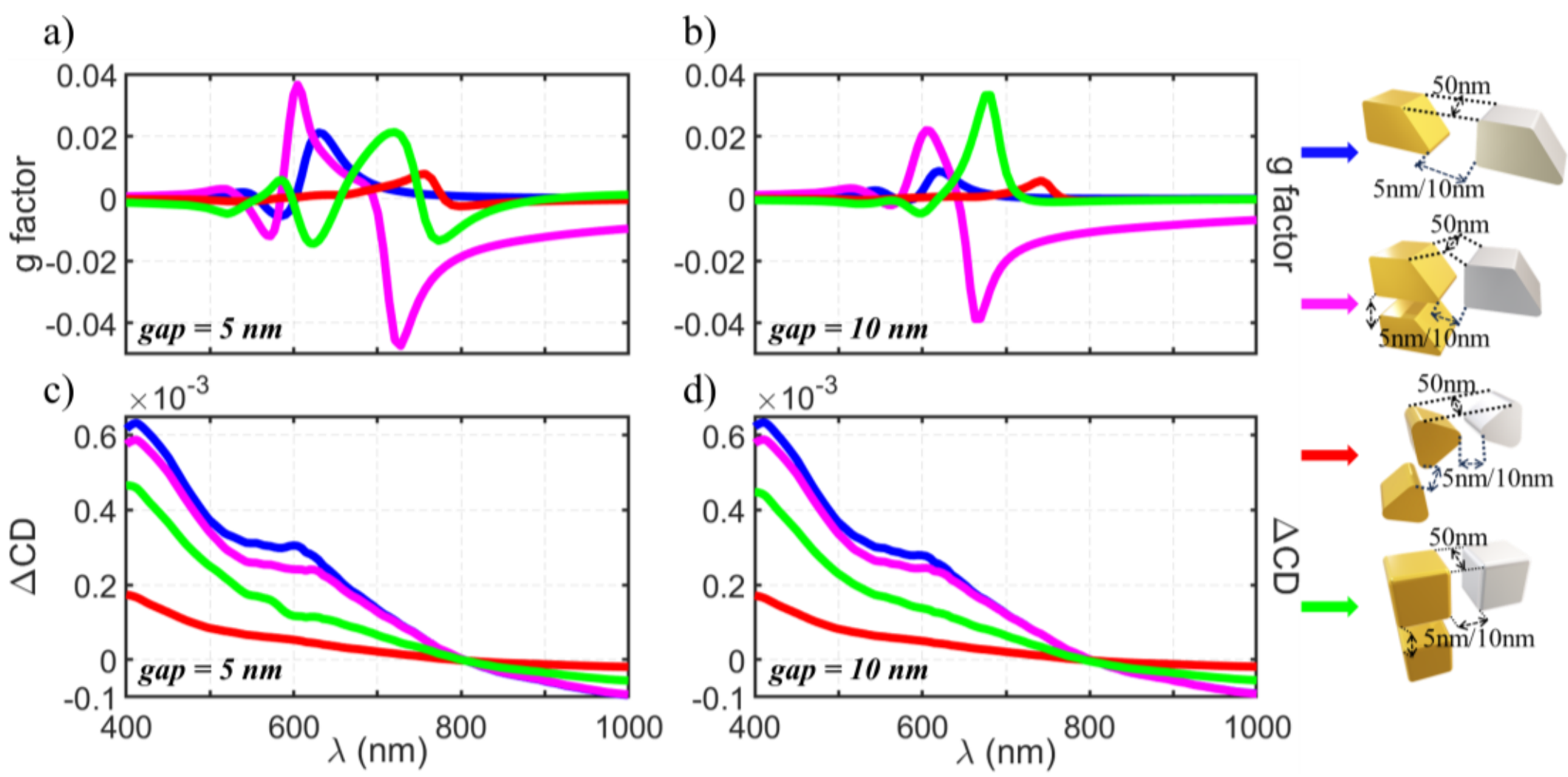

**Fig. 5**. (a.b) *g* factor spectra in the non-magnetized systems, and (c.d) magnetic tunability of CD spectra, as *MCD* = Δ*CD*, calculated from FEM simulations for bimetallic chiral structures with 50 nm thicknesses, and nanoparticle gaps of 5 nm (a.c), and 10 nm (b.d), for dimer of Au Ni trapezoids (blue curves), trimer of Au, Au and Ni trapezoids (magneta curves), trimer of Au, Au, and Ni triangles (red curves), and trimer of Au, Au, and Ni cubes (green curves).

Fig. 5 illustrates the significant effect of shape, number of elements, and distances of the nanoantenna elements of these chiral magneto-plasmonic metasurfaces on their chiro-optical activity, quantified by *g* factor, and MO activity, quantified by MCD signal (see Figs. S8-S12 of the Supporting Information for examples of the electric near-field). Increasing the interparticle gap in hybrid arrays of closely packed nanoparticles from 5 nm to 10 nm is expected to weaken the particle–particle coupling and resonances, giving rise to hybridized plasmon modes from near-field coupling between the closely spaced nanoparticles which enables significant enhancement and engineering of the chiro-optical effects. However, this assumption does not hold for the trimer of cubes. The *g* factor spectra of the cubic trimer show a sharper resonance for 10 nm gaps between nanoparticles, compared to the system with 5 nm gaps (green curves in Fig. 5a.b). Indeed, with decreasing interparticle gap, stronger plasmonic coupling is evident from the large bands in transmittance (dips); see Supporting Information, Figs. S15-S16 for further spectra.

Consistent with the results shown in Fig. 4c, d, increasing the gaps between nanoparticles in any of the studied bimetallic systems does not significantly impact the tunability of the CD spectra (Fig. 5c, d). The dimer of Au and Ni trapezoids exhibits the highest tunability among the studied hybrid nanoantennas (blue curves, Fig. 5). However, the *g* factor and CD magnitude of this system are lower with broader resonance compared to the trimer of trapezoids and trimer of cubes (Fig.5, and Fig. S14–S15).

To achieve enhanced results, or to observe different resonances in a desired frequency range, these parameters could be further tailored to obtain optimal structures with suitable response (results considering different heights and inter-nanoparticle distances are given in the Supporting Information Figs. S13–17S).

## 3.3. Comparison of CD and MCD of the proposed systems

In order to compare the results obtained previously, we show in Fig. 6 the CD and ΔCD signals to describe the magnetic tunability curves for all the studied magneto-chiral structures, including variations in height ($h$ = 40 nm and $h$ = 50 nm), and spacing between the nanoparticles ($d$ = 5 nm and $d$ = 10 nm) in the case of bimetallic nanoantenna systems (Fig. 6a-c correspond to structures with 50 nm thickness, and Fig. 6b-d correspond to structures with 40 nm thicknesses).

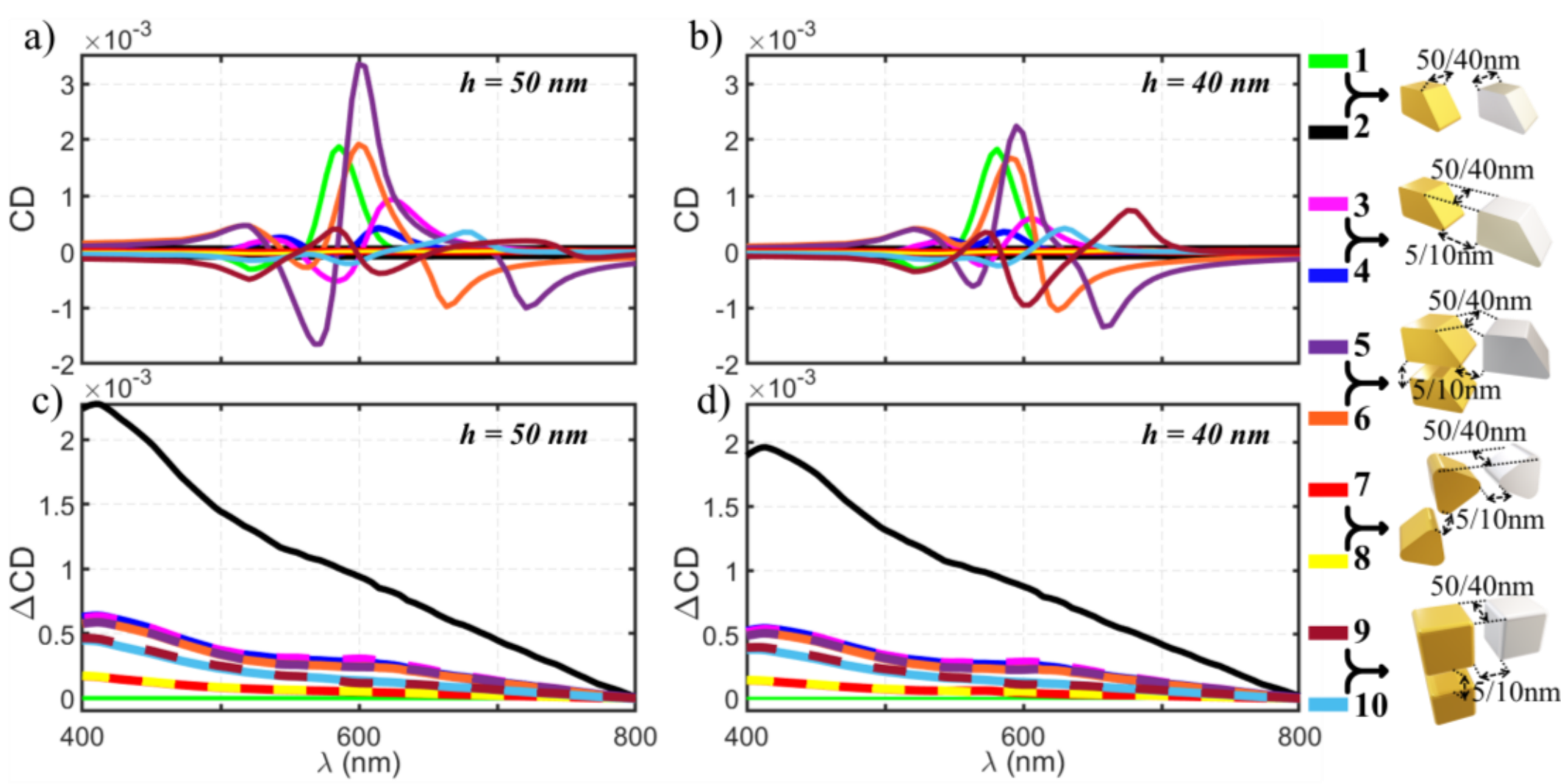

**Fig. 6**. The CD spectra of the non-magnetized magneto-chiroptical metasurfaces and magnetic tunability of the CD spectra of the magneto-chiroptical metasurfaces in the presence of external magnetic field applied in opposite directions (±*H*), as *MCD* = Δ*CD*. Nanoparticles in the metasurfaces have different thicknesses, either 50 nm ($h$ = 50 nm) in height (a,c), or 40 nm ($h$ = 40 nm) in heights (b,d), and different inter-nanoparticle distances ($d$ = 5 nm, or $d$ = 10 nm). For a single Au (1) or Ni (2) nanotrapezoid; dimer of Au and Ni trapezoids with (3) $d$ = 5 nm, (4) $d$ = 10 nm; trimer of Au, Au, and Ni trapezoids with (5) $d$ = 5 nm, (6) $d$ = 10 nm; trimer of Au, Au, Ni triangles with (7) $d$ = 5 nm, and (8) $d$ = 10 nm; and trimer of Au, Au, Ni cubes with (9) $d$ = 5 nm, and (10) $d$ = 10 nm.

It can be inferred from Fig. 6c-d that a single Ni trapezoidal nanoantenna exhibits the highest level of magnetic tunability compared to other proposed structures with a maximum of $MCD_{max} \approx 2.2 \times 10^{-3}$ (Fig. 6c, black lines), and a monomer of purely plasmonic Au trapezoid does not exhibit any magnetic dependency in its differential CD response (Fig. 6c,d, green lines). Interestingly, while the overall magnitude of the CD response is lower for thinner nanoparticles, the CD signal of a monomer of Au trapezoid does not follow the same trend as the rest of the structures (Fig. 6a-b). Unlike other structures that have a lower intensity compared to a monomer of Au trapezoid, the hybrid trimer of Au, Au, and Ni trapezoids is presenting an enhancement on order of ~2-fold in the magnitude of the CD spectra. The main result of Fig. 6 is that the extent of CD and its magnetic tunability follow the structural modifications as a rule, showing that the nanoparticle dimensions are effective parameters in controlling the magnetic adjustability of the chiro-optical response. This means that by changing the size, shape, periodicity, and spacing between nanostructures in hybrid systems, we can control the required magnetic field to modulate the transmission or absorption of light in these systems. This could have applications in developing new types of optical sensors or modulators.

# 4. Discussion

Two optical properties that are central in the design of efficient active chiral magneto-plasmonic systems are the strength of the magnetic modulation and the optical resonance quality [23]. Reaching strong magnetic modulation and tunability without degrading the plasmonic characteristics is a continuous challenge and the practical applications are typically prevented by a magneto-plasmonic tradeoff [23]. Au, a good plasmonic metal, has sharp chiro-optical resonances [26, 65], but no MO response [22, 25] (Fig. 4, green lines). Ni, a ferromagnetic metal, on the other hand has a strong magneto-plasmonic response, but without a clear plasmonic resonance peak [25], that extends to the wavelength range 400-1000 nm that we studied here [25] (Fig. 6 and Fig. S17, black lines).

Thus, combining plasmonic Au elements with MO Ni elements might result in a reduction in the quality of plasmonic, MO, or both characteristics [23]. This trend could indeed be seen in our results for hybrid nanoantennas compared with monomers: A trimer of Au, Au and Ni trapezoids has the highest CD (ca. $CD_{max} \approx 3.3 \times 10^{-3}$) (Fig. 6a, and Fig. S15g), and $g$ factor (ca. $g\ factor_{max} \approx 35 \times 10^{-3}$ and ca. $g\ factor_{min} \approx -47 \times 10^{-3}$) (Fig. 5a, and Fig. S16g), while the trapezoidal Ni monomer has the highest MCD (ca. $2.2 \times 10^{-3}$) (Fig. 6c-d, and Fig. S17).

However, the choice of an optimized nanoantenna can facilitate an optimal chiro-optical MO response. As an example, while the maxima of CD is lower for the dimer of trapezoids (ca. $1 \times 10^{-3}$) and trimer of triangles (ca. $0.03 \times 10^{-3}$) compared to that of the Au monomer (ca. $1.4 \times 10^{-3}$) (Fig. 6a-b), the trimer of hybrid trapezoids is exhibiting an even larger CD response ( $CD_{max} \approx 3.3 \times 10^{-3}$) than any other system (Fig. 6a,b), yet the dimer of Au Ni trapezoids has the highest MCD among the hybrid structures which amounts to ca. $0.63 \times 10^{-3}$ (Fig. 5c,d). Also, our results show that increasing the radius on the tips of the proposed nanoparticles does not have a strong effect on the magnetic tunability of the chiro-optical response, while it has a clear effect on the CD, and *g* factor signal (CD, *g* factor, and MCD results considering different radius of the nanoparticle tips are shown in the Supporting Information, Fig. S15-S17, respectively),

# 5. Conclusion

Our work systematically studies the optical behavior of organized metamaterials that experience both time-reversal and space-inversion symmetry breaking, as a result of magnetism, and chirality, respectively [66, 67]. We have presented chiral plasmonic nanoantennas made of one or more elements which accommodate the magnetic modulation of the chiral absorption of light. The incorporation of a magnetized Ni nanoparticle as the active magneto-plasmonic element allows for the magnetic modulation of the near- and far-field chiro-optical response in these systems.

Among the proposed structures, the maximum tunable parameter of calculated MCD pertains to a single Ni nanotrapezoid and amounts to ca. $2.2 \times 10^{-3}$ in the wavelength range of 400 to 420 nm. Within this wavelength range, an external magnetic field notably enhances both the CD and *g* factor maxima, exhibiting ~2 orders of increase in the magnitude compared to the non-magnetic system, highlighting the external magnetic field's substantial effect (Fig. 3d). Although the obtained tunable parameter for the proposed structures is not very large, it is conceivable that larger values of tunability could be achieved by implementing changes in the nanoantennas design. It should be noted that the maximum magnetic tunability value of the chiro-optical response achieved in 2D magneto-plasmonic [22, 26] and chiro-optical plasmonic systems [68] is about 0.2 to 0.3 degrees. The available technology makes it possible to design highly sensitive chiral MO setups capable of detecting $\Delta A$ signals in the range of $10^{-6}$- $10^{-7}$(which would be around $3 \times 10^{-6}$ mdeg of ellipticity) [50].

The results of this study indicate that tunability of a chiral magneto-plasmonic nanoantenna can be controlled by changing, among others, the periodicity, height and the inter-nanoparticles distances. This control over the tunability of the chiro-optical response could have important implications for the development of new optical devices with tunable chiro-optical properties. Another finding of this paper which could be of high value to experimentalists is that even though increasing the size of the fillet used as the radius on the tips of the nanoparticle reduces the intensity of the CD signal (Fig. S15) and the *g* factor (Fig. S16), it does not have a strong impact on the differential magnetic tunability of the chiro-optical response (Supporting Information, Fig. S17).

Our findings not only validate the fact that all the proposed structures are chiral, but also demonstrate the possibility to magnetically control the chiro-optical response in these systems. In contrast to isolated Au-Ni pairs of magneto-plasmonic nanodiscs, where the overall chiro-optical response is zero in the absence of the magnetic field and the applied magnetic field induces only a weak chiral response purely due to the magnetic nature of the Ni element [22], our numerical simulations reveal that in addition to magnetic modulation of the chiro-optical response, monomers and dimers of nanotrapezoids exhibit enhanced chiro-optical performance, which is not observed in single or paired nanodiscs.

The essential tuning element is the MOA nanoparticle of the nanoantennas, while its near-field coupling to the rest of the elements provides the desired changes in the overall far-field chiro-optical response. The bimetallic antenna accommodates a large structural flexibility, and a whole palette of strongly chiral MO metasurfaces can be produced by slightly modifying the configuration attributes of the system with nanofabrication.

This proof of concept opens the route to new chiro-optical sensing techniques by designing chiral magneto-plasmonic structures with increased local optical chirality, and magnetic field modulations and modifications of chiro-optical effects.

**Acknowledgement**

This work was supported in part by the Waldemar von Frenckell foundation and the University of Sistan and Baluchestan. KK acknowledges support from the Svenska Kulturfonden (Arvid och Greta Olins fond, grant No. 200129). The funders and supporters had no role in study design, data collection and analysis, decision to publish, interpretation of the data; preparation review and approval of the manuscript; and decision to submit the manuscript for publication.

The computer resources of the Finnish IT Center for Science (CSC) and the FGCI project (Finland) are acknowledged. KK gratefully acknowledge the hospitality of DIPC.

**Contributions**

All authors contributed to the writing and revision of the manuscript for intellectual content.

Supporting Information

# Chiro-Optical Structures with Magnetizable Plasmonic Elements for Modulating the Chiral Transmission of Light

Kaysiyavash Kaykavoosi[1, 3, *], Nicklas Anttu[1], Mario Zapata-Herrera[2], Javad Ahmadi-Shokouh[3], Hamideh Dashti[3], Majid Rashidi Huyeh[4]

[1] Department of Physics, Faculty of Science and Engineering, Åbo Akademi University, FI-20500 Turku, Finland

[2] Donostia International Physics Center (DIPC), Donostia-San Sebastián 20018, Spain

[3] Department of Communications Engineering, Faculty of Electrical & Computer Engineering, University of Sistan and Baluchestan, Zahedan, Iran

[4] Department of Physics, Faculty of Basic Sciences, University of Sistan and Baluchestan, Zahedan, Iran

[*] Corresponding Author: Kaysiyavash Kaykavoosi, kaysiyavash.kaykavoosi@abo.fi

## S1. Simulation details

Realization of a practical plasmonic antenna for near-field MO measurements is subjected to the separation of the plasmonic element from the magnetic material. The complexity of a plasmonic antenna interacting with a magnetic material is such that traditional mathematical approaches are not viable for solving the light-scattering problem with the Maxwell equations, especially for antenna structures of non-trivial geometries [69]. To optically describe the MO effects in our systems, we used the finite element method (FEM), implemented in the commercial software COMSOL Multiphysics® [48], mainly due to (1) the ability of this software in allowing the user to define the permittivity as a complex non-Hermitian tensor, and (2) the ability for adaptive mesh-refinement to effectively capture the curvature and edges of the nanostructures.

The simulated system consisted of several domains, as depicted in Fig. S1. Floquet periodic boundary conditions were imposed in the transverse directions on the side walls to periodically repeat the unit cell in the *xy*-plane. Since perfectly matched layers (PMLs) absorb any light scattered into them, they were placed at the two remaining ends of the geometry (in the *z* direction, with the negative *z* direction toward the particles from the top side, with the substrate on the bottom side), that in effect continue the system to infinity. The physical domain is the domain limited by the two PMLs and the periodic boundary conditions. A circularly polarized pulse is emitted from the top boundary of the physical domain from an active input port in the negative *z* direction towards the particles. The wave equations in the input port for CPL is set in the following way:

$$LCP = \begin{cases} E_x = {}^{1}\!/\!{\sqrt{2}} \\ E_y = {}^{+j}\!/\!{\sqrt{2}} \end{cases}, \quad RCP = \begin{cases} E_x = {}^{1}\!/\!{\sqrt{2}} \\ E_y = {}^{-j}\!/\!{\sqrt{2}} \end{cases}. \tag{S1}$$

The nanoantenna with finite substrate is surrounded by air. Poynting vector was then integrated above and below the nanoparticles to yield information of reflectance ($R$) and transmittance ($T$), respectively. From these findings, the absorptance of the nanoparticles can be obtained as $A = 1 - R - T$. From the simulations, we obtain also the electric field distribution throughout the system, and as a numerical control, we also calculated the absorption in each nanoparticle through a volume integration of the spatially resolved absorption. Dimension of the unit cell with the largest unit cell parameter of 120 nm was selected to avoid lattice resonances in the lateral directions. The refractive index of the substrate was $n_{\mathrm{sub}} = 1.5$, mimicking the optical properties of Pyrex.

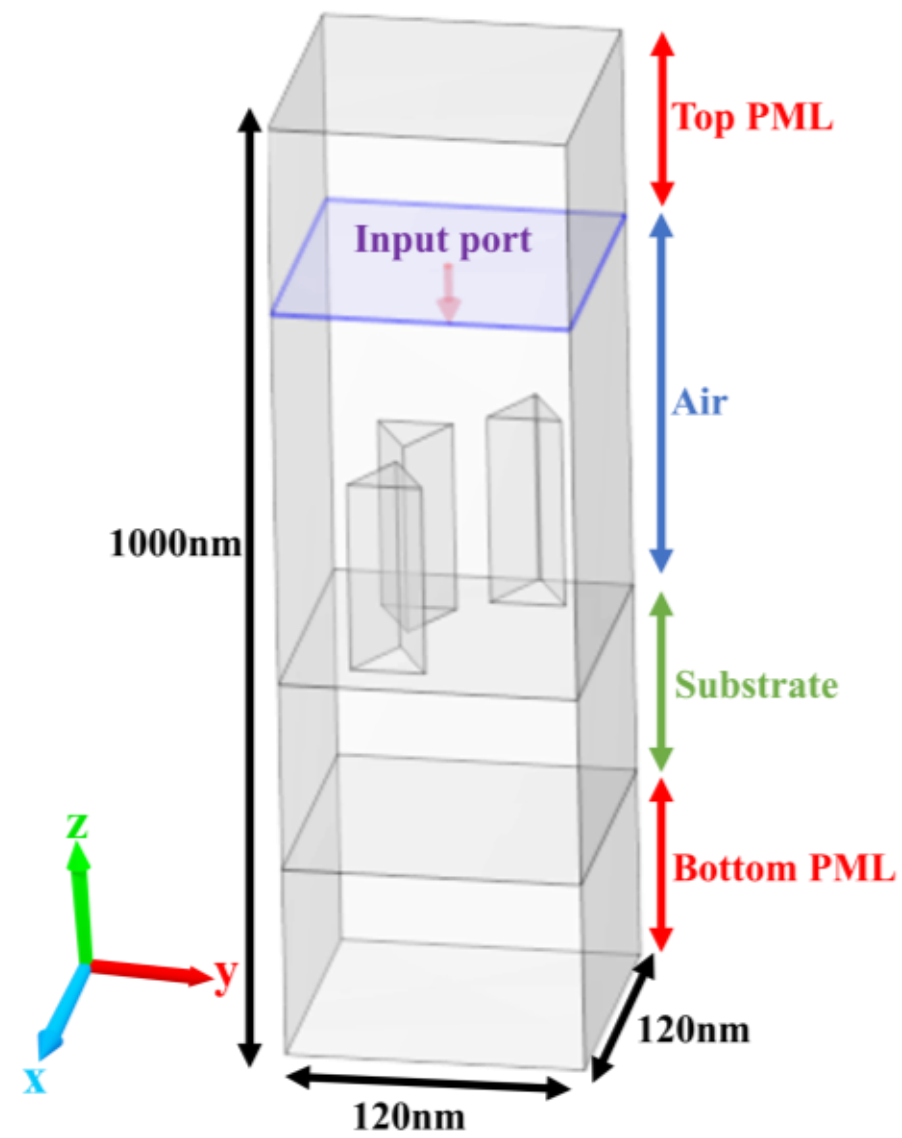

**Fig. S1.** Visualization of the numerical domain used to calculate the CD and MCD signals of the chiral magneto-plasmonic structures (note that the figure is not to scale in the in-plane vs. out-of-plane directions).

Tetrahedral elements were used to mesh all domain with the maximum mesh element size below $\lambda_{min}/20$, where $\lambda_{min}$ = 400 nm is the shortest wavelength considered for the incident field. For the mesh elements corresponding to nanoparticle domains, the size was ten times finer than the largest element size in the nanoparticle domains and on the range of 1 nm to 2 nm, which were verified to properly resolve the considered structures.

## S2. Optical and magneto-optical properties of metals

The permittivity tensor, $\overleftrightarrow{\varepsilon}$, contains the optical information of any material in the linear, local, and time-harmonic response that does not show direct magnetic response at the optical frequency, that is, such that the permeability tensor at the optical frequency is given by $\overleftrightarrow{\mu} = \mu_0$ with $\mu_0$ the (scalar) permeability of vacuum [17]. In isotropic and non-magnetic materials, this tensor is diagonal [27, 45]. For the permittivity values of Au and Ni (non-magnetic), we used the wavelength dependent dielectric optical functions based on the values measured in experiments [70, 71]. In isotropic and non-magnetic materials, this tensor is diagonal [27]:

$$\overleftrightarrow{\varepsilon} = \varepsilon_0 \begin{pmatrix} \varepsilon & 0 & 0 \\ 0 & \varepsilon & 0 \\ 0 & 0 & \varepsilon \end{pmatrix}. \qquad \text{(S2)}$$

where $\varepsilon$ corresponds to $\varepsilon = n^2$, $n$ is the refractive index of the material, and $\varepsilon_0$ is the permittivity of vacuum [45].

Applying an external magnetic field generates anisotropy (magnetization) in ferromagnetic nanoparticles (Ni in our study), which is added to the simulations by introducing the off-diagonal elements in their permittivity tensor. This anisotropy in MO materials can be controlled, utilizing an external magneto-static ($H_{DC}$) field. When the magneto-static field is applied along the *z*-axis, also called the polar configuration, the permittivity tensor will have the following form [28, 53]:

$$\overleftrightarrow{\varepsilon} = \varepsilon_0 \begin{pmatrix} \varepsilon & iQ & 0 \\ -iQ & \varepsilon & 0 \\ 0 & 0 & \varepsilon \end{pmatrix}. \tag{S3}$$

Here, $Q$ is called the complex Voight coefficient (or the MO constant), which is a function of magnetization, created by an externally applied magnetic field, and $\varepsilon_0$ is the permittivity of vacuum [28, 45, 53]. The MO constant is proportional to $H_{DC}$. A key principle in the field of magneto-optics is that if the orientation of $H_{DC}$ is reversed, then $Q(-H_{DC}) = -Q(H_{DC})$, as long as for $H_{DC} = 0$, the off-diagonal components of the permittivity tensor are zero [28, 55]. MCD is the difference in the CD signal for a system magnetized in opposite directions [26] (keeping in mind the configuration presented in Fig. 2). The calculated magnetic tunability of the chiro-optical response ($MCD = \Delta CD$) depends on the relative orientation of the applied magnetic field ($\pm H$), leading to an aligned magnetization of the magnetic sample in the same direction ($\pm M$) [26, 36, 66, 72].

## S3. Using fillets for the tips of the nanoparticles

Due to the challenges and limitations associated with fabrication of perfectly sharp tips, structures with rounded tips can be more realistic when taking fabrication limitations into account since nanostructures often have a finite radius of curvature there [26, 73]. We therefore considered four different cases of tips for the structure: Structures with sharp corners, and structures with rounded tips, that is, fillets, with a radius of 1 nm, 2 nm, and 3 nm. These variations provide a more comprehensive understanding of the behavior and properties of these nanostructures; and how the corners affect the chiro-optical response. Additionally, whether with or without fillets, the center-to-center distances in dimer and trimer structures remain constant. Consequently, the transition from sharp to rounded tips, along with reduction in the surface area resulting from implementation of fillets, simultaneously affect the inter-nanoparticle distances, that is, tip-to-tip distances. In the main text, we used the results acquired for structures with 2 nm radius on tips.

Introducing fillets as radius of curvature on nanoparticle tips leads to subtle changes in the inter-nanoparticle distances when the center-to-center distance is kept fixed. Fig. S2 illustrates the geometric parameters of interest, that is, radius of curvature on the tips of each nanoparticle and distance between each pair of nanoparticles. The effects of these parameters could be used to optimize the chiro-optical properties of nanoantennas.

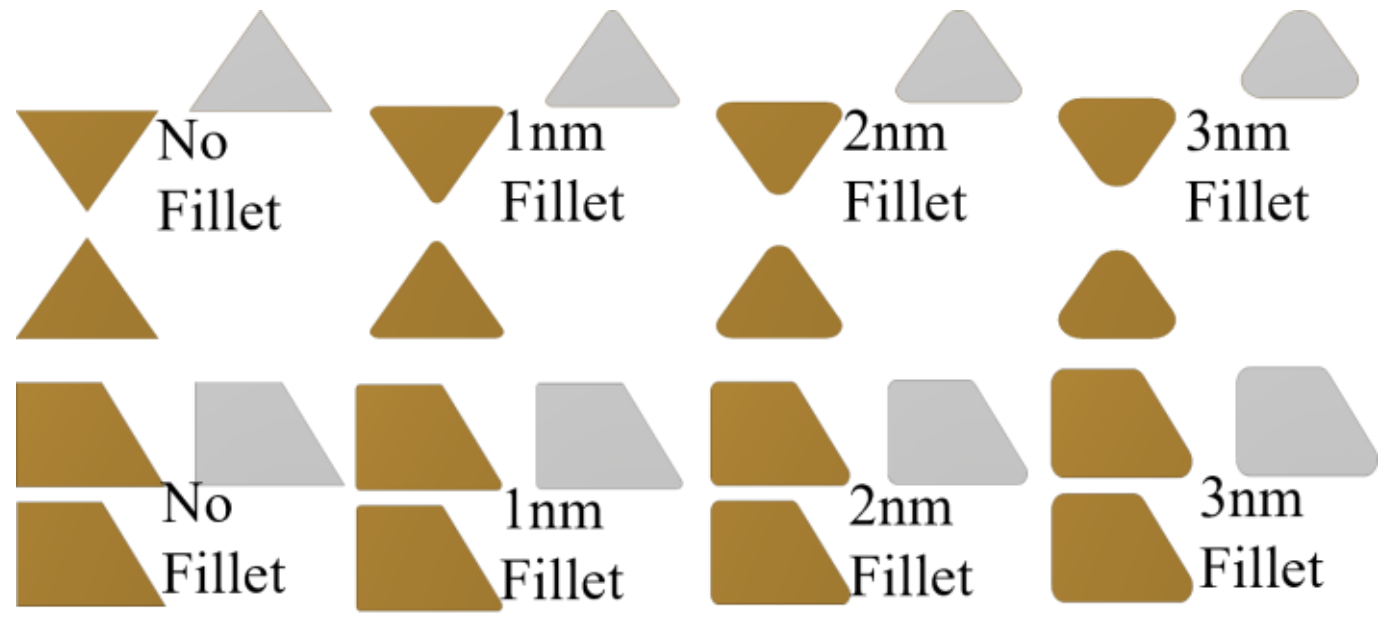

**Fig. S2.** Configuration of trimer of triangles and trapezoids without fillets, with 1 nm, 2 nm, and 3 nm curve radius on corners of each nanoparticle.

In the following, for all supplementary results presented in Figures S3-S30, the results for structures with no fillets, as well as with 1 nm, 2 nm, and 3 nm fillets are included for clarity and completeness. A summary of the effect of the fillets is given in Section S8. We allow for a rounding of the nanoparticle tips, quantified by a fillet that specifies the radius of curvature on the tips (see Supplementary Section S.3 for details). Importantly, we have seen that the magnetic tunability of the CD signal does not appear to be strongly dependent on the choice of the fillet, at least in the range of 1-3 nm fillets (Supporting Information, Fig. S17). For the results presented in the main text, we chose the fillet to 2 nm.

## S4. Effects of fillet on results

When fillets are added to tips of the structures, the gap distances between the nanoparticles increases proportionally with the radius of the fillets. To investigate the impact of maintaining constant inter-nanoparticle gaps while incorporating fillets on nanoparticle tips, we will compare the results for a trimer of triangles with 1 nm fillets. In the first scenario, as shown in Fig. S3a, the inter-nanoparticle gaps are 5 nm or 10 nm before the addition of fillets, and the inclusion of fillets results in a slight increase in these nanoparticle distances, while the center-to-center distances between nanoparticles are kept constant (results in Fig. S4, solid lines). In the second scenario, as shown in Fig. S3b, the inter-nanoparticle gaps are kept constant at 5 nm or 10 nm after fillet incorporation (results in Fig. S4, dashed lines).

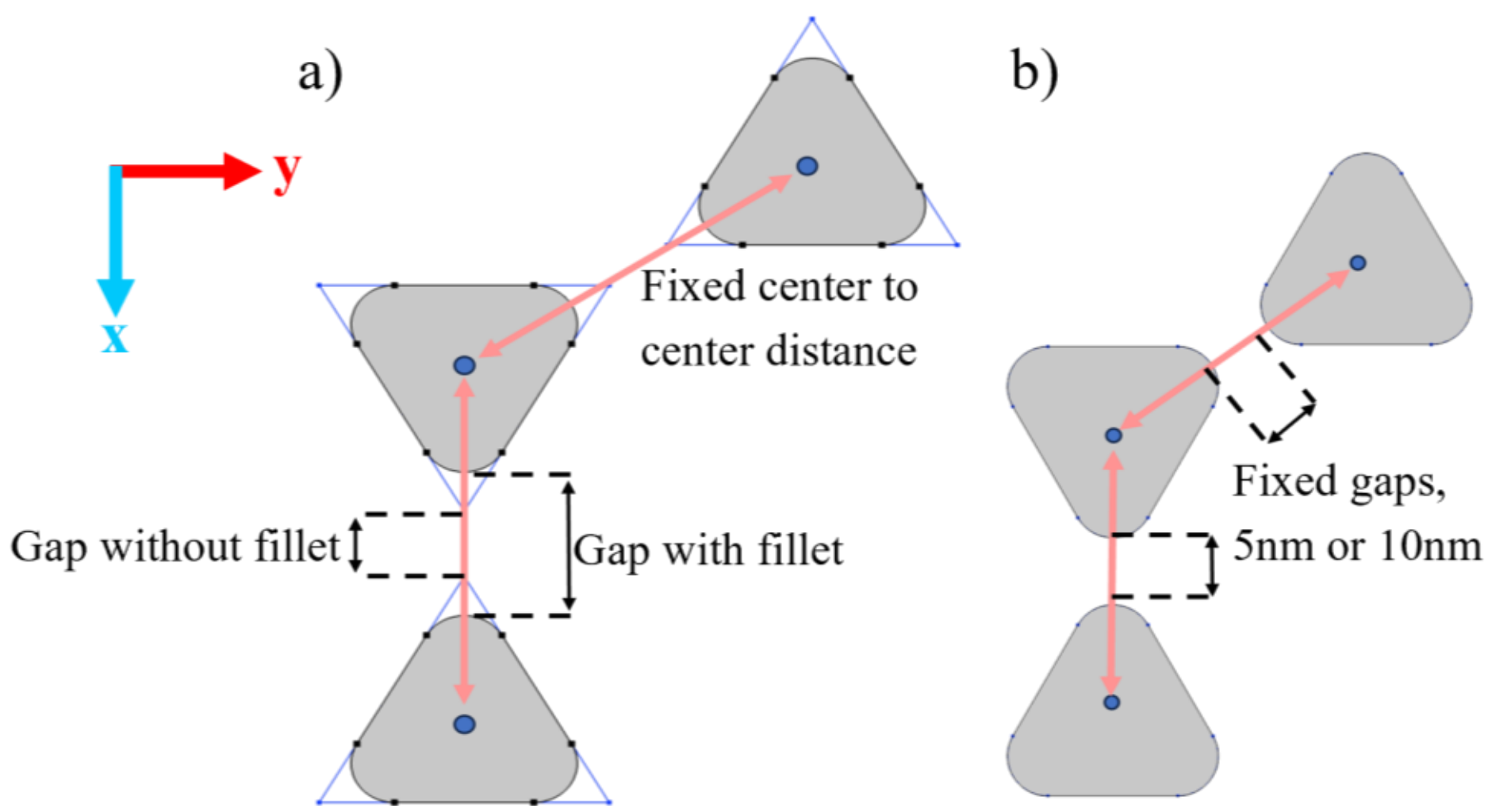

**Fig. S3.** Configuration of trimer of chiral magneto-plasmonic triangles with 1 nm curve radius on tips, with constant center-to-center distances, with and without fillets (a), and with fixed gaps after the addition of fillets (b).

The CD and *g* factor spectra are presented in Fig. S4 for the triangle trimer nanoantennas with constant center-to-center distances but unfixed gaps (similar to the structure in Fig. S3a), as well as for fixed gaps after the addition of fillets (similar to the structure in Fig. S3b), with different gap distances and heights.

For ease of comparison, we also present both the CD and *g* factor signals of this structure with fixed gaps (Fig. S5a, and Fig. S5c), and with gaps not being fixed but with a constant center-to-center distance (Fig. S5b, and Fig. S5d), respectively.

The transmission spectra of these two systems can be seen in Fig. S6, which is plotted with respect to keeping the height of the structures fixed (Fig. S6a, and Fig. S6b), keeping the gaps fixed, similar to the system in Fig. S3b (Fig. S6c), and similar to the system in Fig. S3a (Fig. S6d).

The results show that keeping the separation gaps constant does not have any strong quantitative impact on our simulated CD and *g* factor compared to when we kept the center-to-center distances constant.

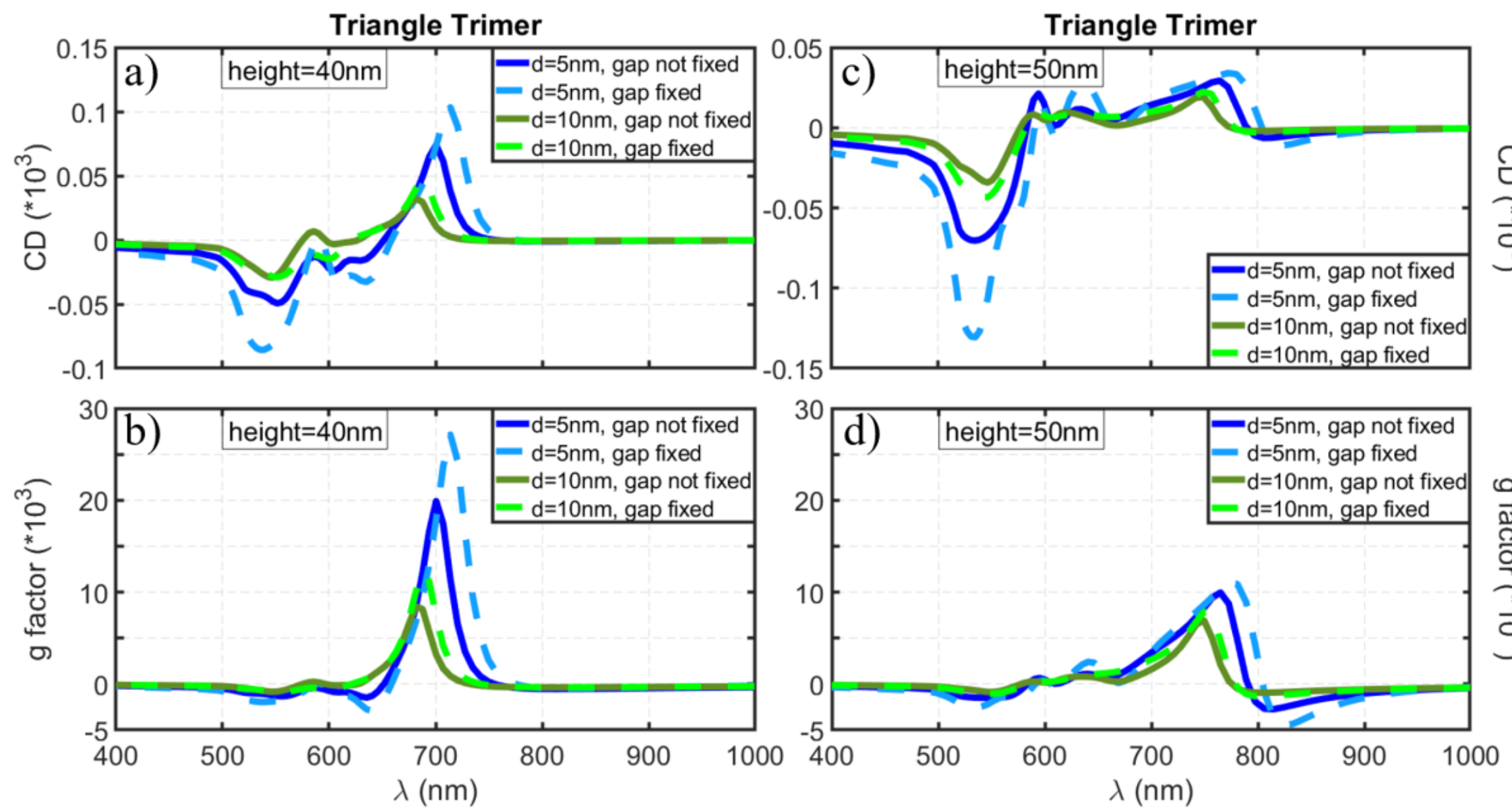

**Fig. S4.** Comparison of CD and *g* factor for the systems presented in Fig. S3 based on nanoparticle thicknesses; (a) CD spectrum, and (b) *g* factor for the system with nanoparticles of 40 nm thicknesses. (c) CD spectrum, and (d) *g* factor for the system with nanoparticles of 50 nm thicknesses.

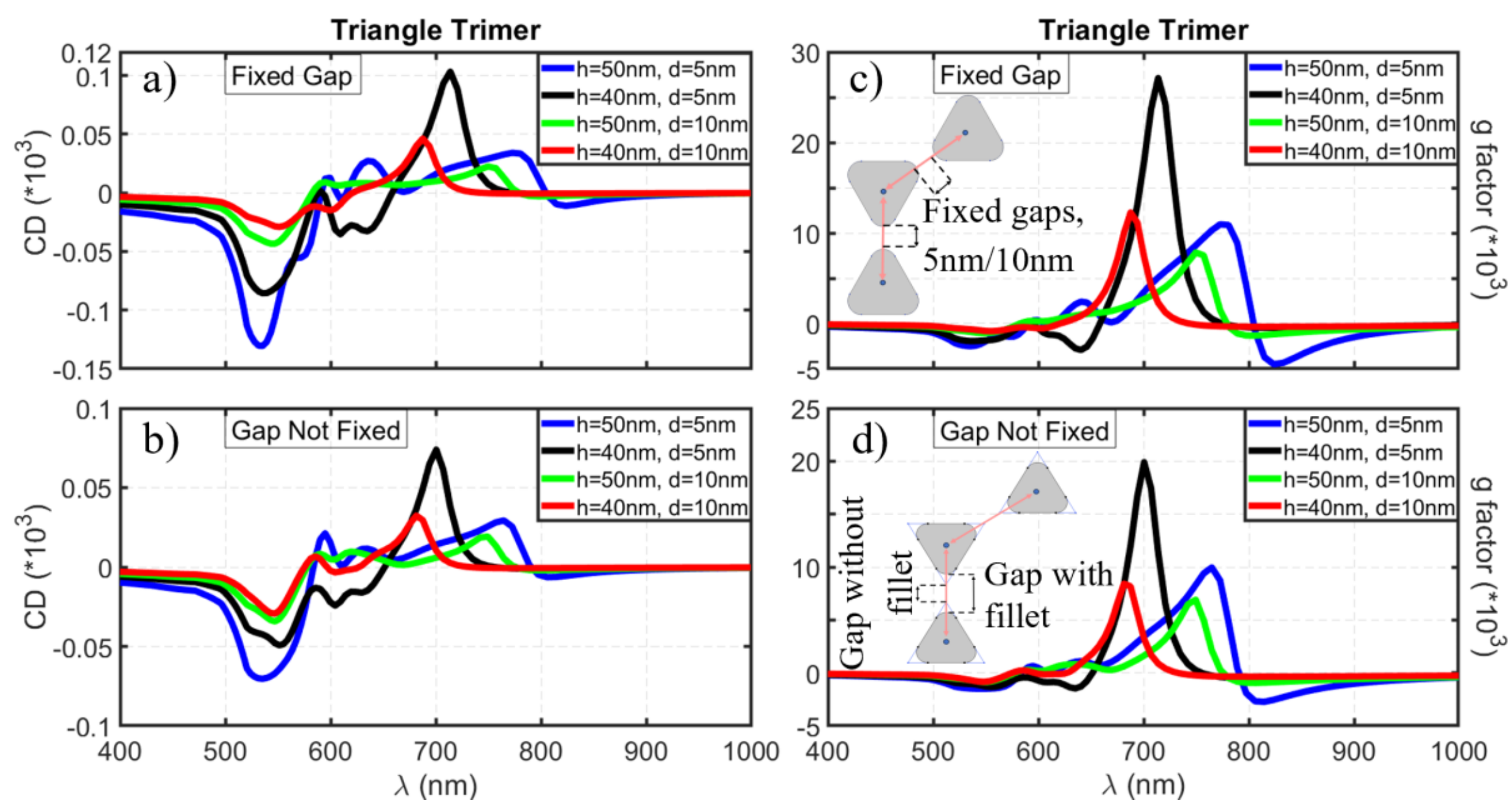

**Fig. S5.** Comparison of CD and *g* factor for the systems presented in Fig. S3 with different heights of $h$ = 40 nm and $h$= 50 nm, and different gap distances of $d$ = 5 nm and $d$ = 10 nm. (a) CD spectrum, and (c) *g* factor for the system presented in Fig. S3b (with fixed gaps). (b) CD, and (d) *g* factor for the system presented in Fig. S3a (with gaps not fixed).

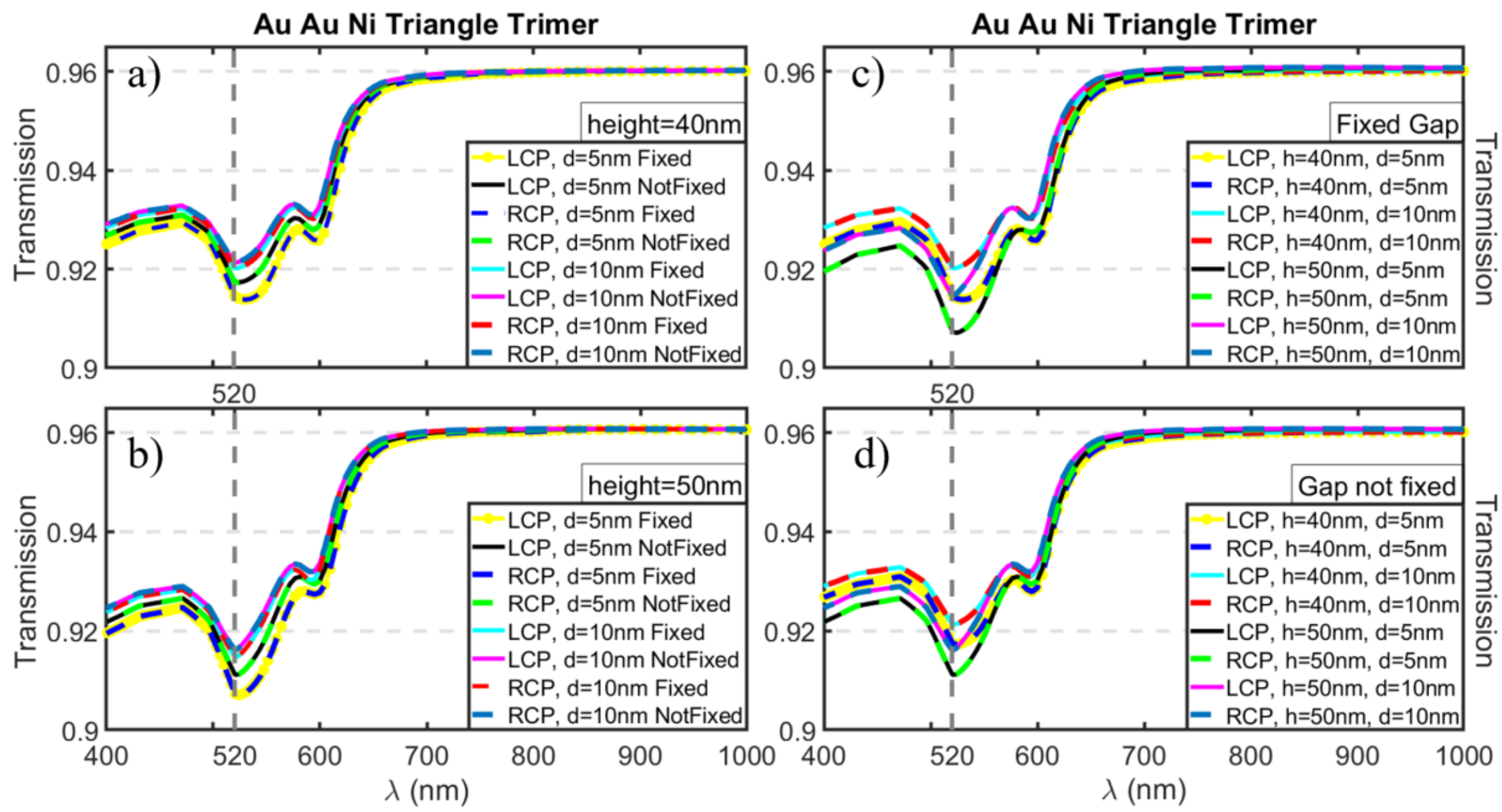

**Fig. S6.** Comparison of transmission spectra for the systems presented in Fig. S3 with different heights of $h$ = 40 nm and $h$ = 50 nm, and different gap distances of $d$ = 5 nm and $d$ = 10 nm; for (a) trimer of nanotriangles with 40 nm height, (b) nanotriangles with 50 nm height, (c) for the system presented in Fig. S3b with fixed gaps, and (d) for the system presented in Fig. S3a with gaps not fixed.

## S4. Transmission spectra and near-field electric distribution in chiral magneto-plasmonic nanoantennas

In Fig. S3, Fig. S4, Fig. S5, and Fig. S6, we present the far-field transmission of RCP and LCP light for the proposed chiral magneto-plasmonic nanoantennas, while taking into account the addition of different curvature radius on nanoparticle tips. In the case of bimetallic systems in Figs. S4-S6, the inter-nanoparticle gaps are either 5 nm or 10 nm. The thickness of the structures also varies and is either 40 nm or 50 nm. In the transmission plots, the dashed vertical lines point out the wavelengths which were used for the results of electric near-field responses.

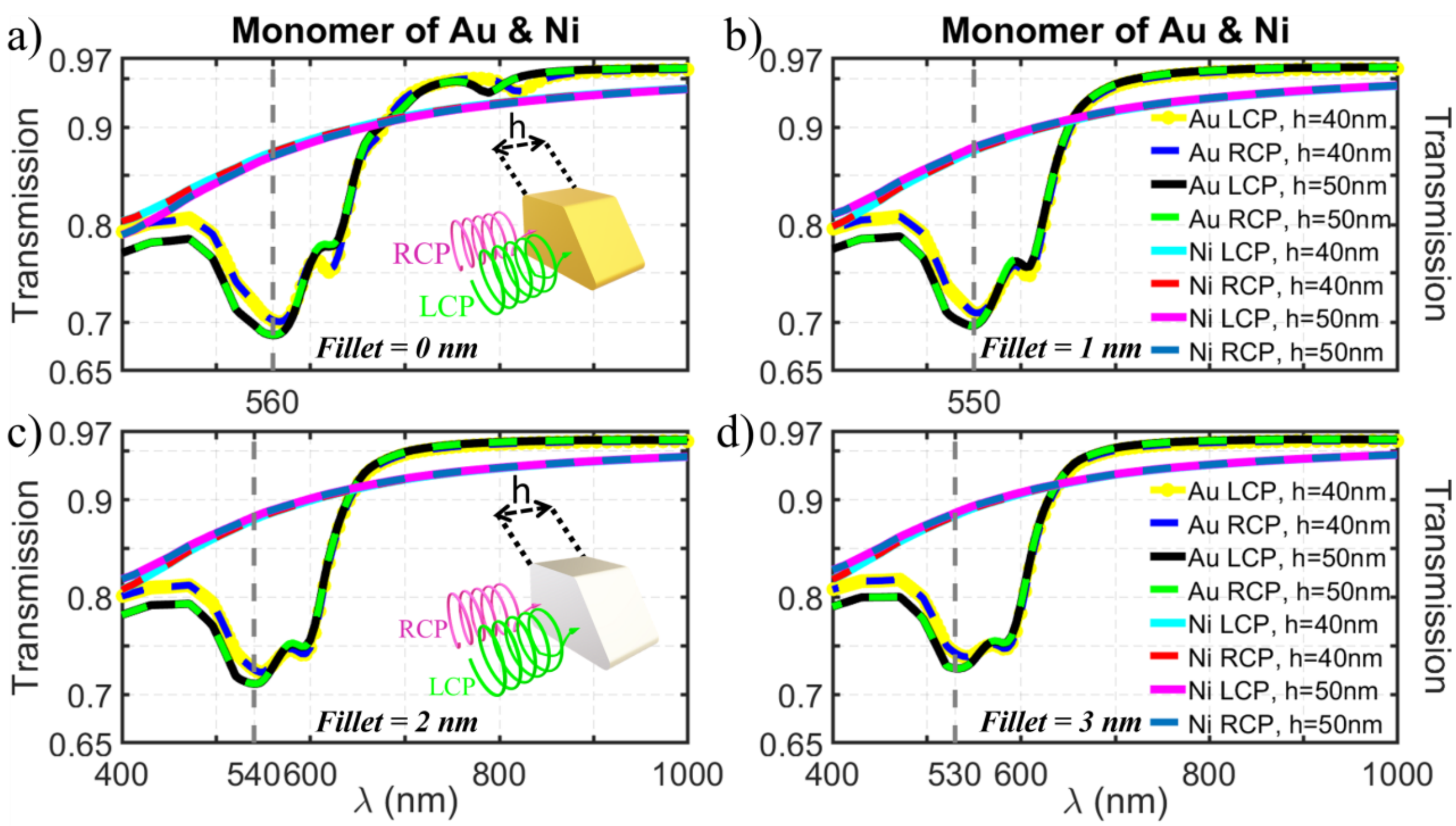

**Fig. S7.** FEM-calculated transmission spectra of LCP and RCP light by chiral nanoparticles of Au and Ni monomers with different thicknesses. The legends show the corresponding CPL handedness as LCP and RCP, and the corresponding structure height in each system as $h$ = 40 nm or $h$ = 50 nm. Far-field transmission of a single Au or Ni trapezoid, (a) without fillets, (b) with 1 nm fillets, (c) 2 nm fillets, and (d) 3 nm fillets on nanoparticle tips.

Strong amplification of the evanescent electric field close to the resonating plasmonic nanoparticles is a characteristic optical effect of plasmonic nanoparticles. In general, the calculated chiro-optical absorption of a chiral nanoantenna results in a distinctly different near-field plasmonic response to the LCP and RCP light. To obtain the chiro-optical properties of the structures, the simulations include applying CPL, and examining the excited electric field in the *xy*-plane with a top-view of the nanoparticles in the visible and near-IR spectral range. The following figures (Figs. S8-S12) are the corresponding results of the electric near-field profile of the chiral nanoantennas illuminated with CPL for varying size of fillets used on the nanoparticle tips, as obtained from FEM simulations. The electric field distributions (shown here normalized to the incoming field and on a logarithmic scale) are non-superimposable for different handedness of CPL, regardless of the radius of curvature on the nanoparticle tips, and the wavelengths are chosen based on the far-field transmission spectra. Keep in mind that even in completely achiral systems, local interactions between the incident and scattered fields (or between light and nanostructures) can produce complex field distributions, i.e. localized regions of chiral near-field responses, but geometrical chirality is necessary for far-field chiro-optical responses to occur [4].

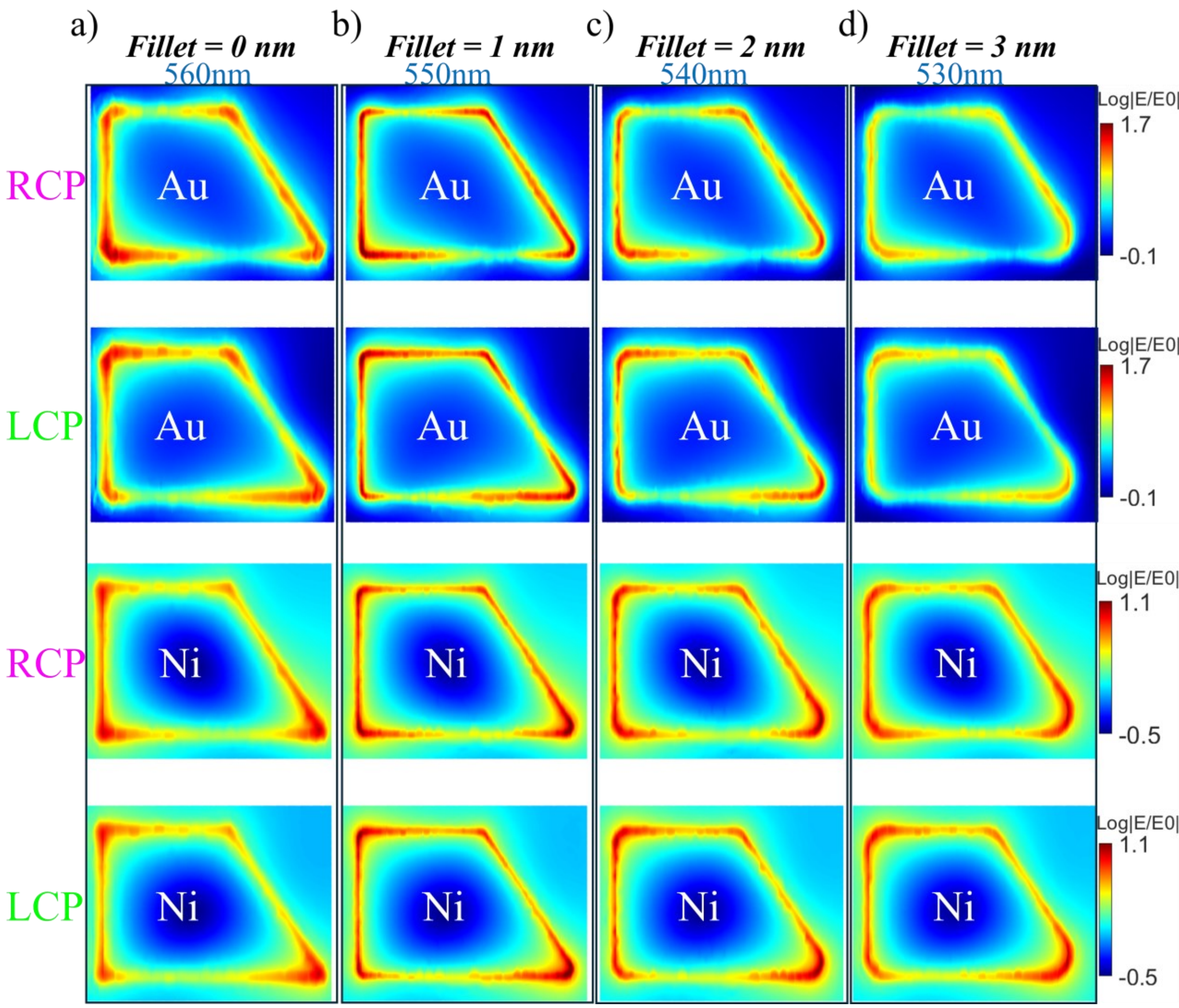

**Fig. S8.** Near-field distribution of the Au and Ni trapezoidal monomers with 50 nm thickness, illuminated with RCP (top row), and LCP (bottom row) light; (a) first column, without fillets at 560 nm, (b) second column, with 1 nm fillets on tips at 550 nm, (c) third column, with 2 nm fillets at 540 nm, and (d) forth column, with 3 nm radius on tips at 530 nm.

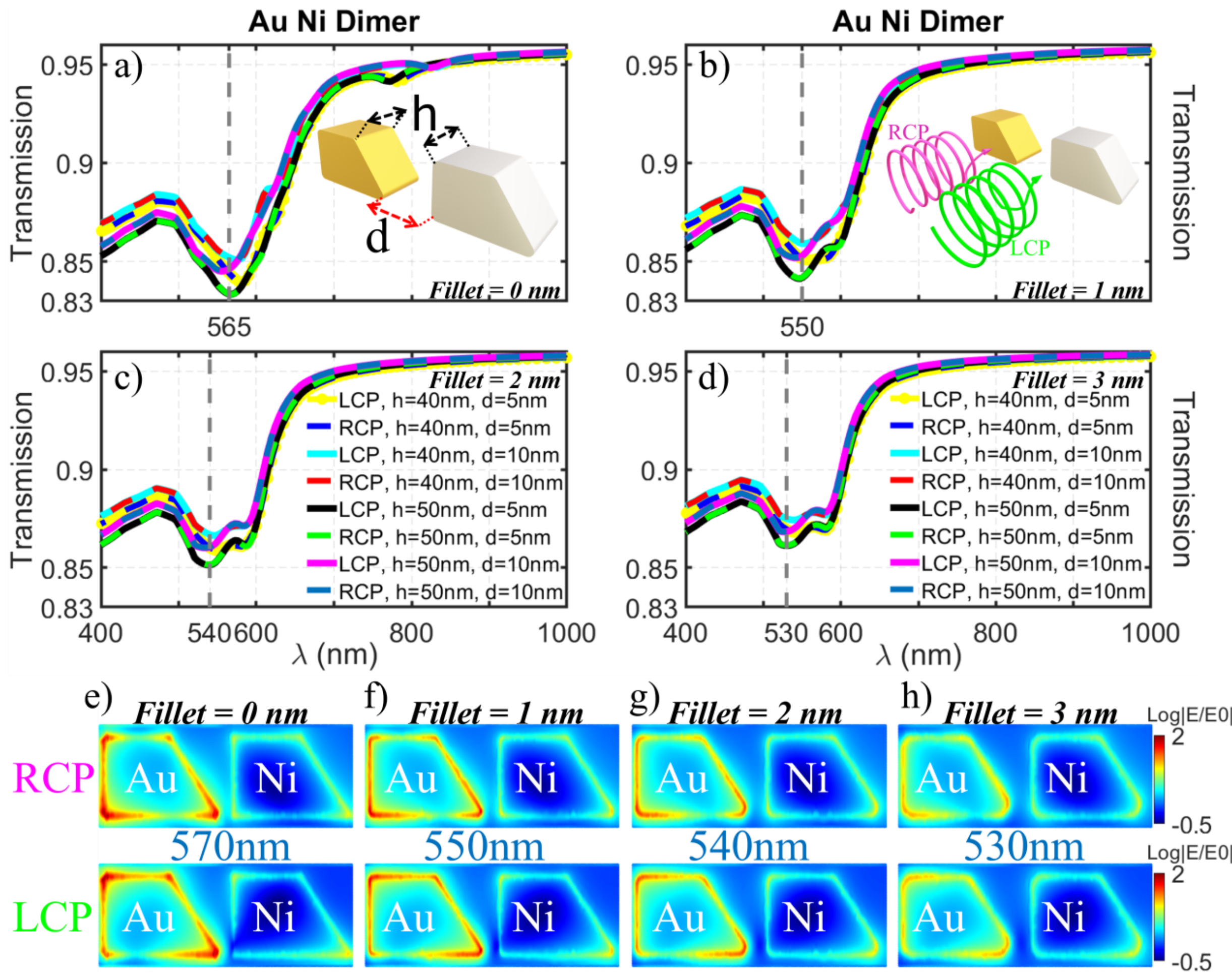

**Fig. S9.** FEM-calculated transmission spectra of LCP and RCP light by Au Ni dimer of chiral trapezoid nanoantennas. The legends show the corresponding CPL handedness (LCP or RCP), the structure thickness as *h* = 40 nm or *h* = 50 nm, and the inter-nanoparticle distances in each system as *d* = 5 nm, or *d* = 10 nm. Far-field transmission of dimer of Au and Ni trapezoidal nanoantennas, (a) without fillets, (b) 1 nm fillets, (c) 2 nm fillets, and (d) 3 nm fillets on nanoparticle tips. Near-field distribution of the Au Ni trapezoidal dimer with 50 nm thicknesses and 5 nm gap distances, illuminated with RCP (top row), and LCP (bottom row) light; (e) without fillets at 570 nm, (f) with 1 nm fillets on tips at 550 nm, (g) 2 nm fillets at 540 nm, and (h) 3 nm radius on tips at 530 nm.

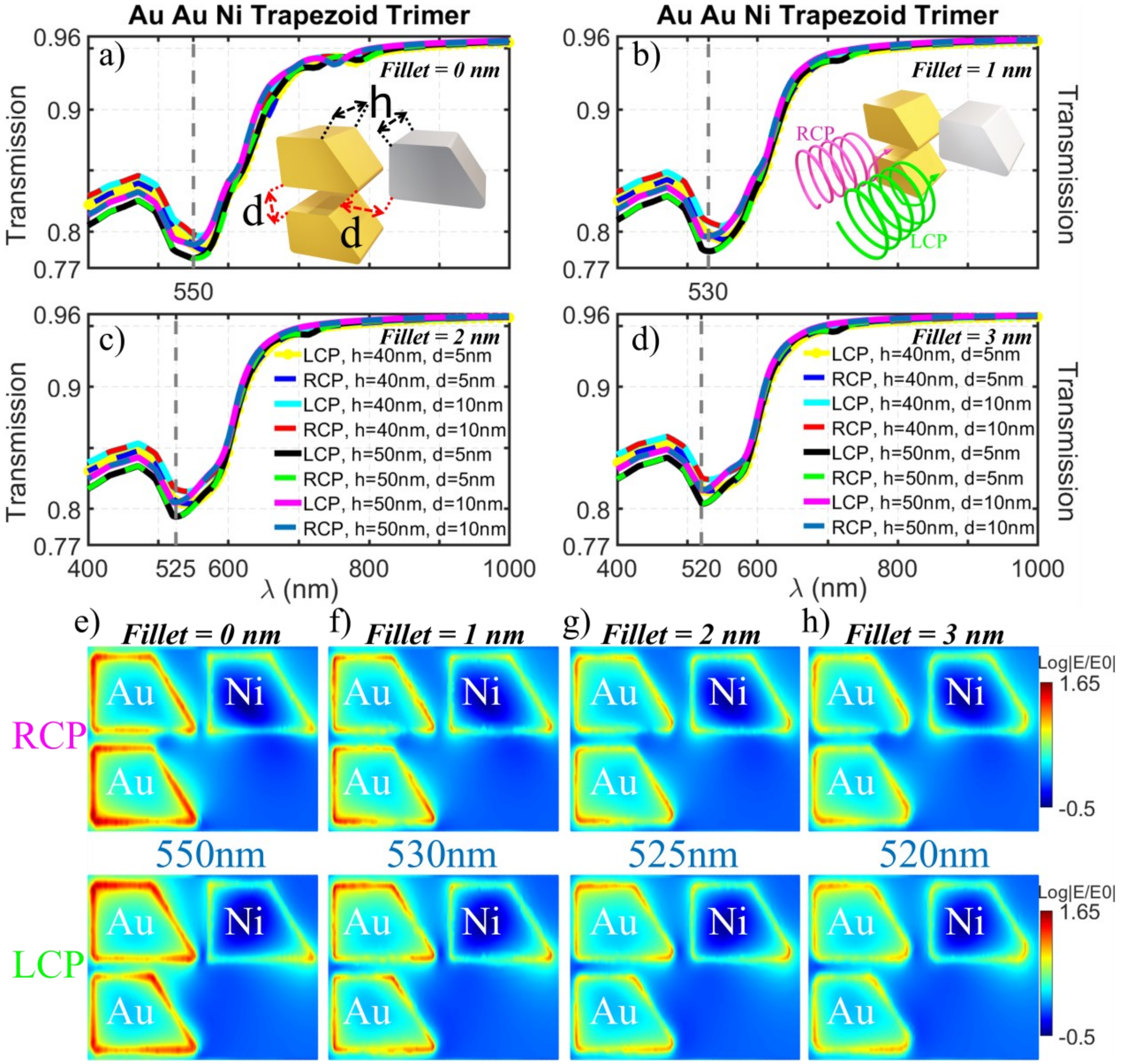

**Fig. S10.** FEM-calculated transmission spectra of LCP and RCP light by Au, Au, and Ni trimer of chiral trapezoid nanoantennas. The legends show the corresponding CPL handedness (LCP or RCP), structure thickness as $h$ = 40 nm or $h$ = 50 nm, and the inter-nanoparticle distances in each system as $d$ = 5 nm, or $d$ = 10 nm. Far-field transmission of trimer of Au, Au, and Ni trapezoidal nanoantennas, (a) without fillets, (b) 1 nm fillets, (c) 2 nm fillets, and (d) 3 nm fillets on nanoparticle tips. Near-field distribution of the Au, Au, and Ni trapezoidal trimer with 50 nm thicknesses and 5 nm gap distances, illuminated with RCP (top row), and LCP (bottom row) light; (e) without fillets at 550 nm, (f) with 1 nm fillets on tips at 530 nm, (g) 2 nm fillets at 525 nm, and (h) 3 nm radius on tips at 520 nm.

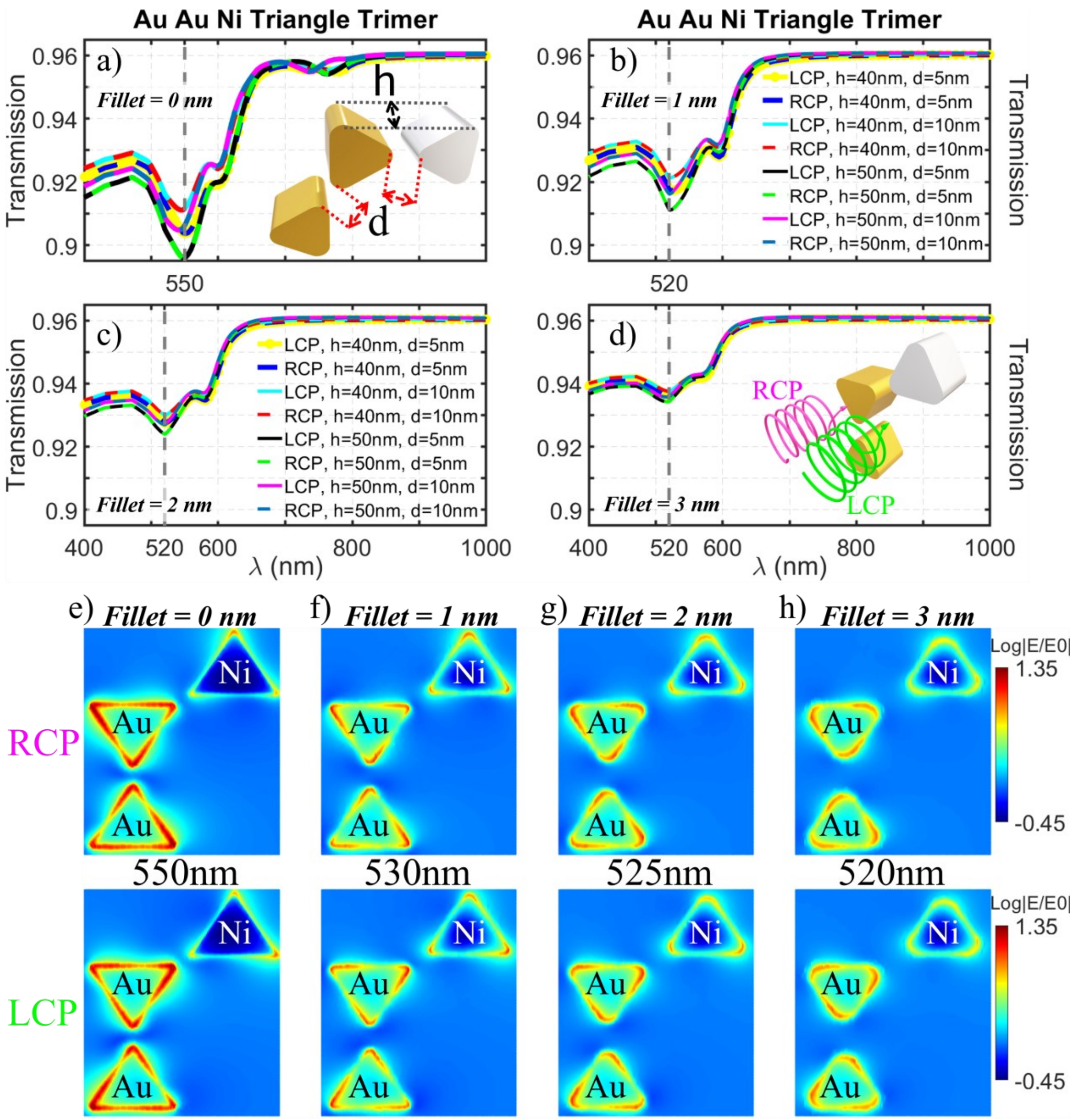

**Fig. S11.** FEM-calculated transmission spectra of LCP and RCP light by Au, Au, and Ni trimer of triangle nanoantennas. The legends show the corresponding CPL handedness (LCP or RCP), the structure thickness as $h$ = 40 nm or $h$ = 50 nm, and the inter-nanoparticle distances in each system as $d$ = 5 nm, or $d$ = 10 nm. Far-field transmission of trimer of Au, Au, and Ni triangular chiral nanoantennas, (a) without fillets, (b) 1 nm fillets, (c) 2 nm fillets, and (d) 3 nm fillets on nanoparticle tips. Near-field distribution of the Au, Au, and Ni triangular trimer with 50 nm thicknesses and 5 nm gap distances, illuminated with RCP (top row), and LCP (bottom row) light; (e) without fillets at 550 nm, (f) with 1 nm fillets at 520 nm, (g) 2 nm fillets at 520 nm, and (h) 3 nm radius on tips at 520 nm.

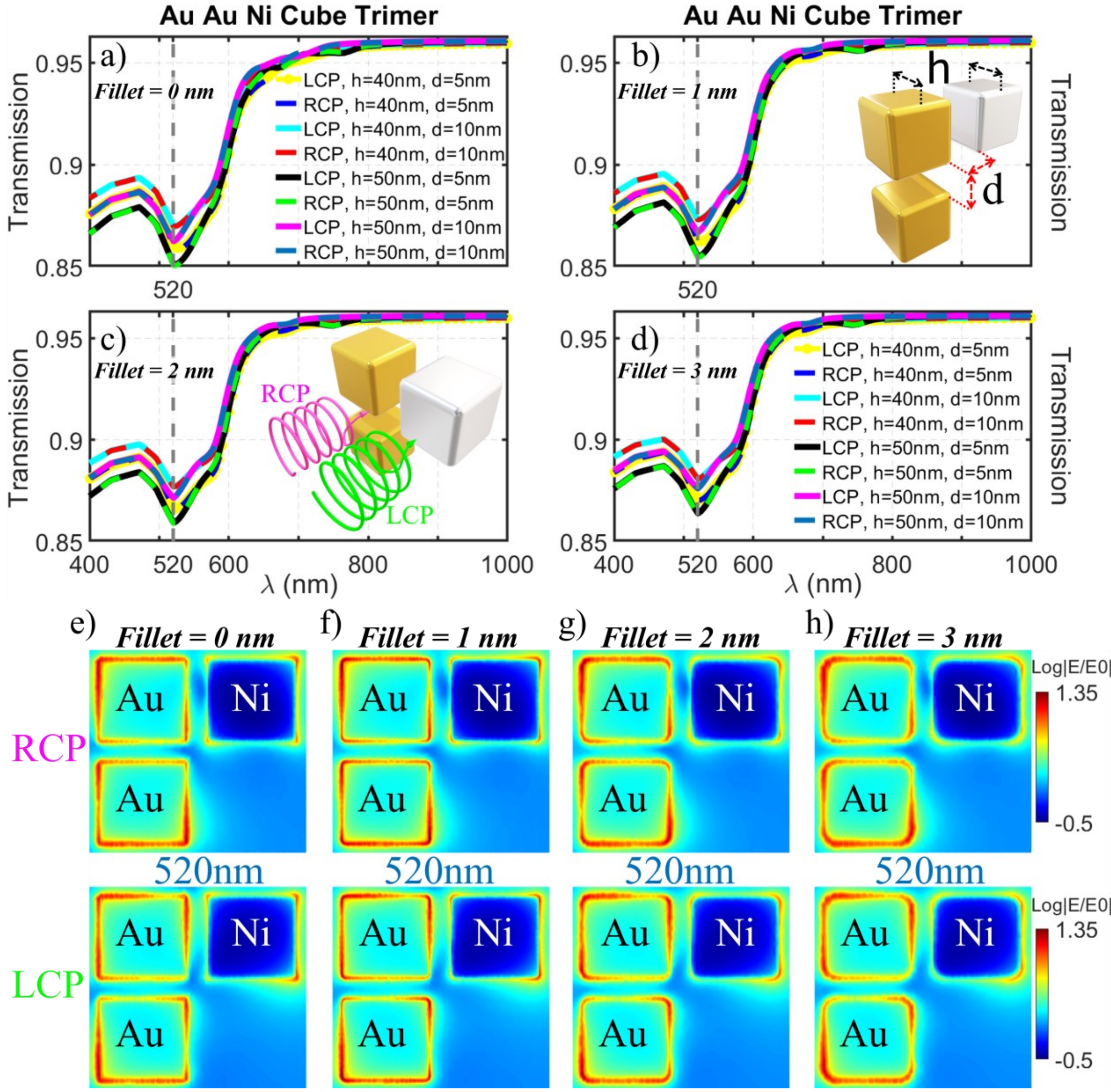

**Fig. S12.** FEM-calculated transmission spectra of LCP and RCP light by Au, Au, and Ni trimer of cubic nanoantennas. The legends show the corresponding CPL handedness (LCP or RCP), the structure thickness as *h* = 40 nm or *h* = 50 nm, and the inter-nanoparticle distances in each system as *d* = 5 nm, or *d* = 10 nm. Far-field transmission of trimer of Au, Au, and Ni cubic chiral nanoantennas, (a) without fillets, (b) 1 nm fillets, (c) 2 nm fillets, and (d) 3 nm fillets on nanoparticle tips. Near-field distribution of the Au, Au, and Ni cubic trimer with 50 nm thicknesses and 5 nm gap distances, illuminated with RCP (top row), and LCP (bottom row) light; (e) without fillets at 550 nm, (f) with 1 nm fillets at 520 nm, (g) 2 nm fillets at 520 nm, and (h) 3 nm radius on tips at 520 nm.

The transmittance spectra of a single Au trapezoid (Fig. S7) and the hybrid systems (dimer and trimer, Figs. S9-S12) indicate that as the thickness of nanoparticles increases from 40 nm to 50 nm, the plasmon band slightly blue shifts, and the transmitted wave weakens.

At a small interparticle distance, the transmittance is naturally low (Figs. S9-S12), but it is mainly the reflectance that exerts the largest influence on the absorptance. Thus, the excitation of the plasmons in an array of closely packed nanoparticles leads to changes in the plasmon relaxation pathways with the radiative damping dominating over non-radiative damping [74]. Consequently, the CD and *g* factor profiles are enhanced for smaller gap distances in all the hybrid systems.

## S5. Effect of dimensions on CD spectra and *g* factor in chiral and magneto-chiral plasmonic nanoantennas

In Fig. 3 of the main text, we presented the CD and the (dissymmetry factor) *g* factor for monomers of Au and Ni trapezoids with thicknesses of 50 nm and a curvature with a radius of 2 nm on the nanoparticle tips, under an externally applied magnetic field in two opposite directions for the Ni monomer nanoantenna. Below, the dimension-dependent CD spectra and the *g* factor are plotted for Au and Ni monomers, and three cases of bimetallic systems in the following figures with different nanoparticle gaps and thicknesses, while taking into account different radius of curvature at the tips, equal to 1 nm, 2 nm, 3 nm, and without rounded tips. The *g* factor is, by definition, within the range from −2 to +2 (See Eq. (1)) [52], and CD data have been reported also as ellipticity ($\theta$), related to the differential absorbance by a factor of 32.98 ($\theta$ (deg) = 32.98 · Δ$A$) [4, 5].

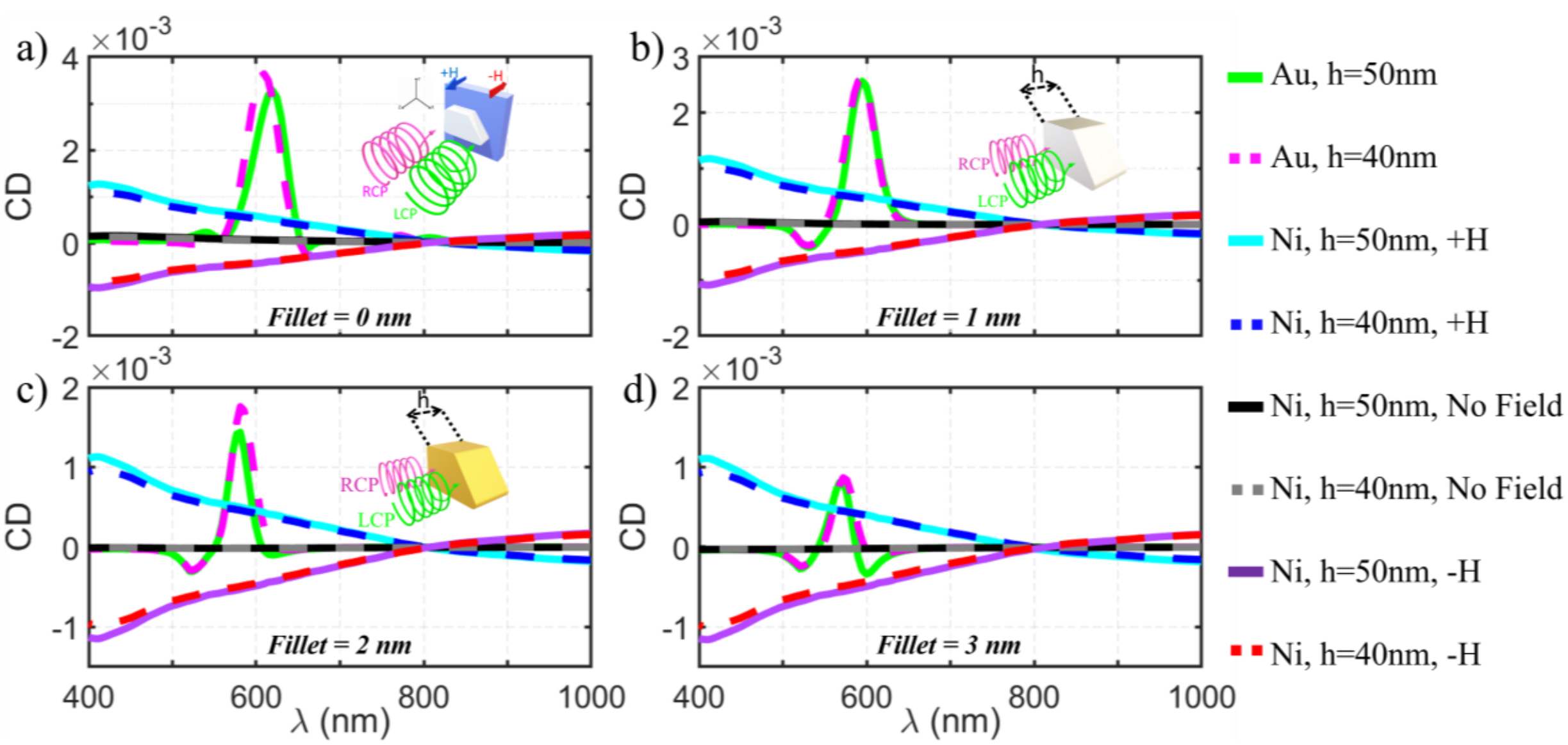

**Fig. S13.** FEM-calculated CD spectra of a single Au trapezoid, and magnetically modulated CD spectra of a chiral magneto-plasmonic monomer of Ni trapezoid illuminated with LCP and RCP light, in the absence of an external magnetic field (No Field), and in the presence of external magnetic field applied in opposite directions (±*H*), structure thicknesses of *h* = 50 nm (solid lines) or *h* = 40 nm (dashed lines), (a) without fillets, (b) with 1 nm fillets, (c) 2 nm fillets, and (d) 3 nm fillets as the curvature radius on tips.

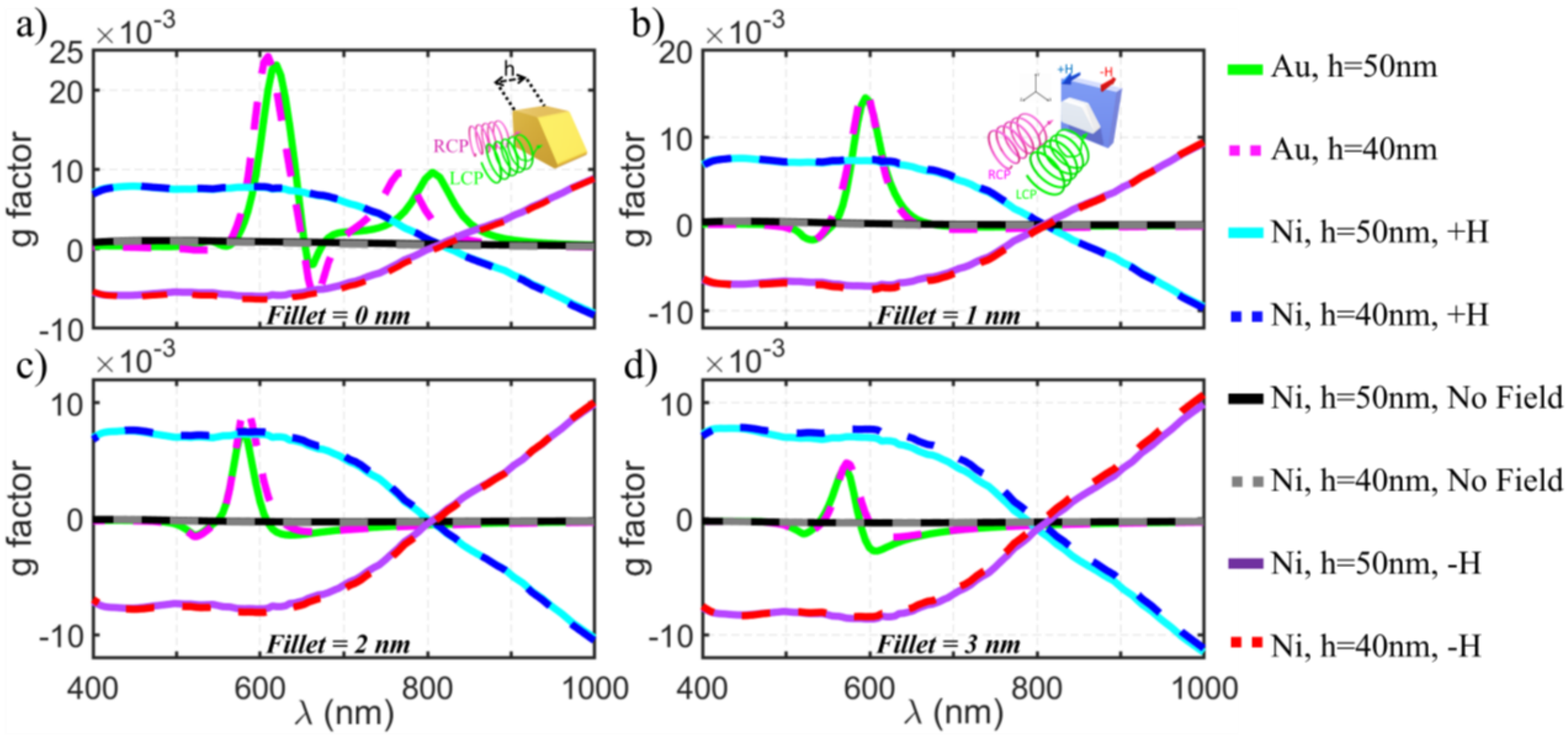

**Fig. S14.** FEM-calculated $g$ factor spectra of a single Au trapezoid, and magnetically modulated $g$ factor spectra of a chiral magneto-plasmonic monomer of Ni trapezoid illuminated with LCP and RCP light, in the absence of an external magnetic field (No Field), and in the presence of external magnetic field applied in opposite directions ($\pm H$), structure thicknesses of $h$ = 50 nm (solid lines) or $h$ = 40 nm (dashed lines), (a) without fillets, (b) with 1 nm fillets, (c) 2 nm fillets, and (d) 3 nm fillets as the curvature radius on tips.

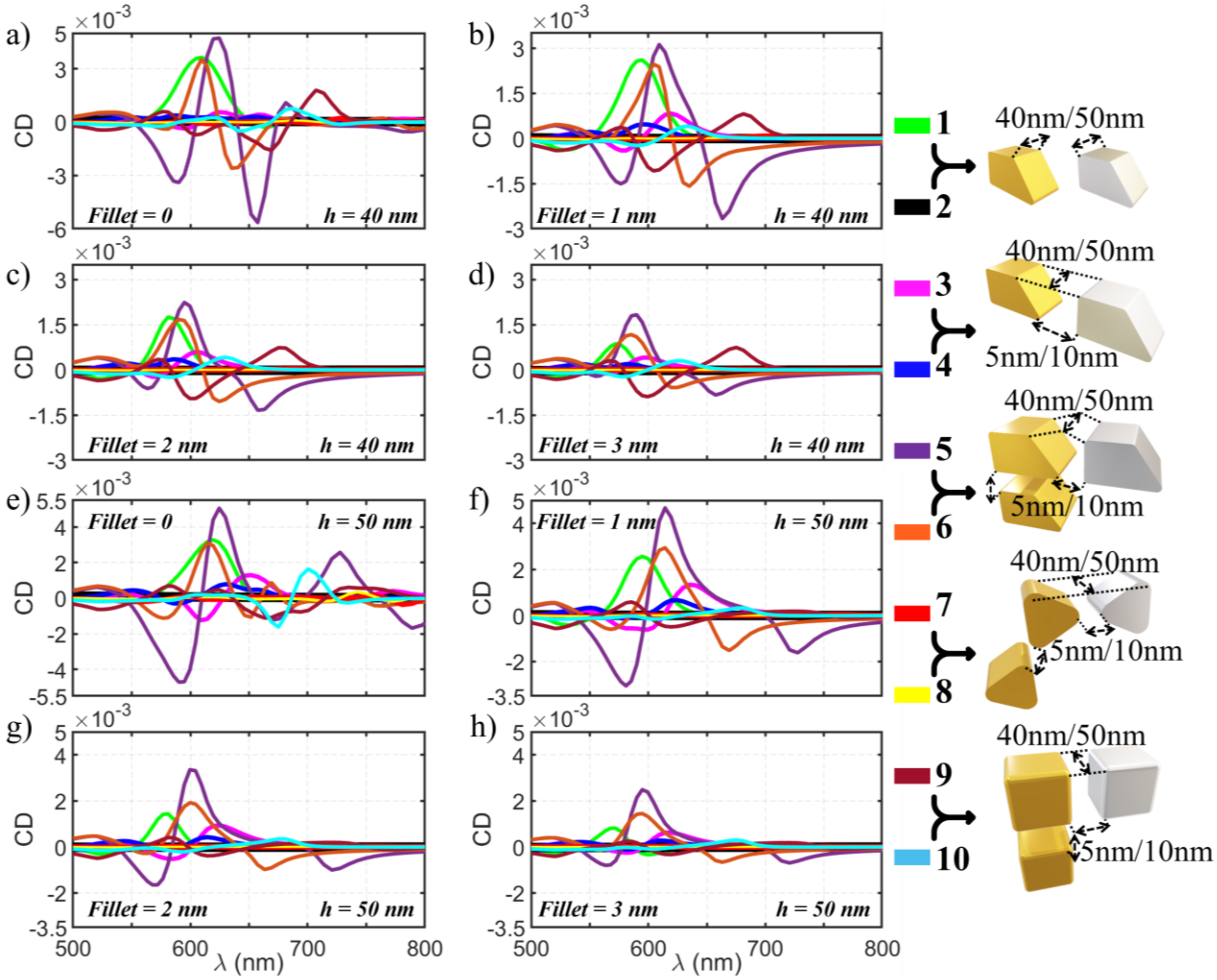

**Fig. S15.** CD spectra of the plasmonic and magneto-plasmonic metasurfaces illuminated with LCP and RCP light, in the absence of an external magnetic field with heights of $h$ = 40 nm (a-d), or $h$ = 50 nm (e-h), with gaps of 5 nm or 10 nm in the case of hybrid systems, (a,e) without fillets, (b,f) with 1 nm fillets, (c,g) 2 nm fillets, and (d,h) 3 nm fillets as the curvature radius on tips. Specifically, the studied systems are: (1) monomer of Au, and (2) monomer of Ni trapezoids. Dimer of Au and Ni trapezoids with (3) 5 nm nanoparticle gaps and, (4) 10 nm gaps. Trimer of Au, Au, and Ni trapezoids with (5) 5 nm nanoparticle gaps and, (6) 10 nm gaps. Trimer of Au, Au, and Ni triangles with (7) 5 nm nanoparticle gaps and, (8) 10 nm gaps. Trimer of Au, Au, and Ni cubes with (9) 5 nm nanoparticle gaps and, (10) 10 nm gaps.

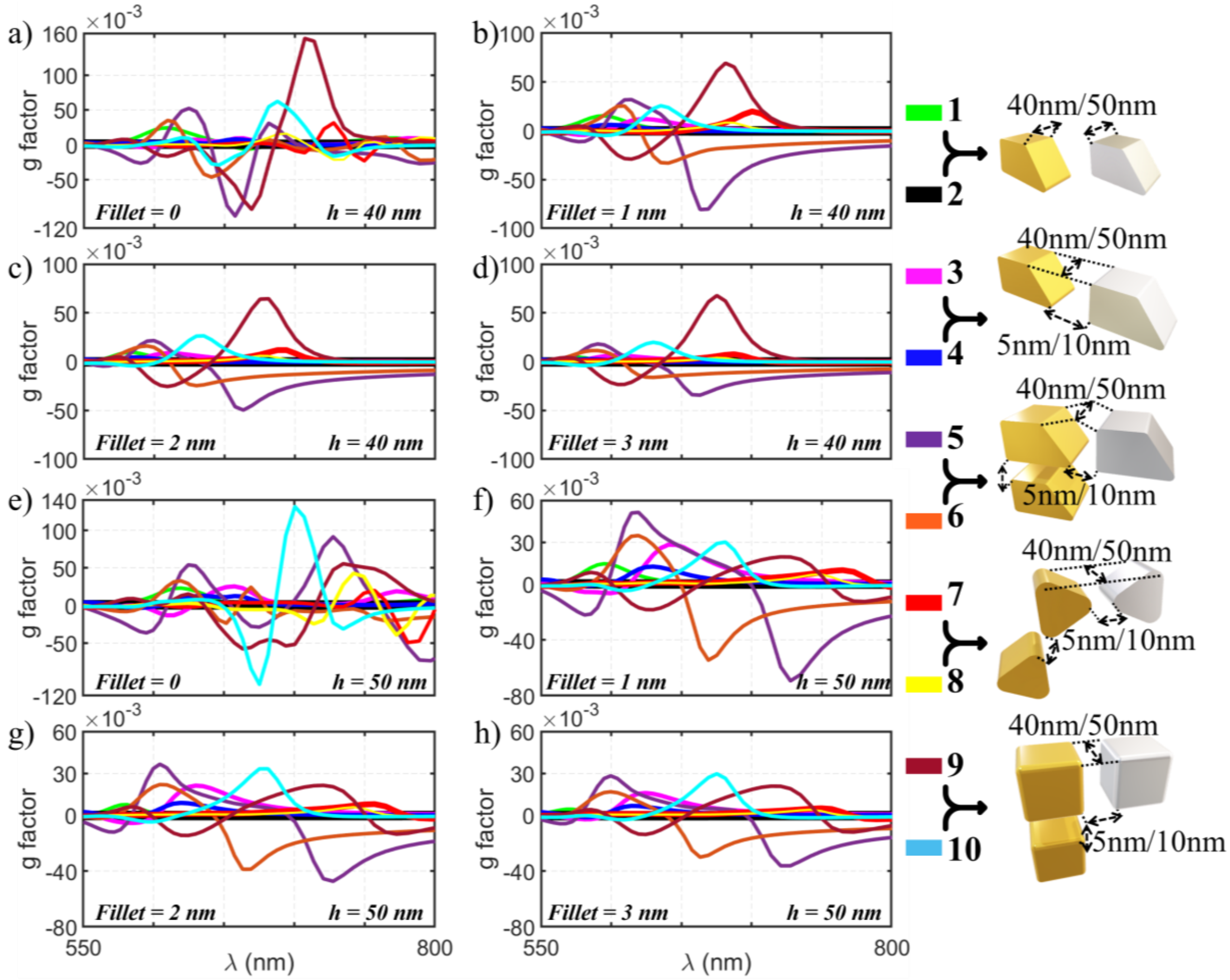

**Fig. S16.** *g* factor spectra of the plasmonic and magneto-plasmonic metasurfaces illuminated with LCP and RCP light, in the absence of an external magnetic field with heights of $h$ = 40 nm (a-d), or $h$ = 50 nm (e-h), with gaps of 5 nm or 10 nm in the case of hybrid systems, (a,e) without fillets, (b,f) with 1 nm fillets, (c,g) 2 nm fillets, and (d,h) 3 nm fillets as the curvature radius on tips. Specifically, the studied systems are: (1) monomer of Au, and (2) monomer of Ni trapezoids. Dimer of Au and Ni trapezoids with (3) 5 nm nanoparticle gaps and, (4) 10 nm gaps. Trimer of Au, Au, and Ni trapezoids with (5) 5 nm nanoparticle gaps and, (6) 10 nm gaps. Trimer of Au, Au, and Ni triangles with (7) 5 nm nanoparticle gaps and, (8) 10 nm gaps. Trimer of Au, Au, and Ni cubes with (9) 5 nm nanoparticle gaps and, (10) 10 nm gaps.

## S7. Magnetic tunability of chiro-optical response

As described in the main text, the choice of Ni as an additional material to Au originates from the fact that Ni, along with, for example Co and Fe, as a ferromagnetic material, has a non-zero magnetization intrinsically. When an external magnetic field is applied, the magnetization aligns with the external field and leads to a difference in the measured CD under opposite magnetic fields, which is known as MCD [26, 38]. In Fig. S30, the results of magnetic tunability of chiro-optical response, MCD (that is, ΔCD), for all the proposed structures and based on the curvature radius on the tips of the nanoparticles are presented.

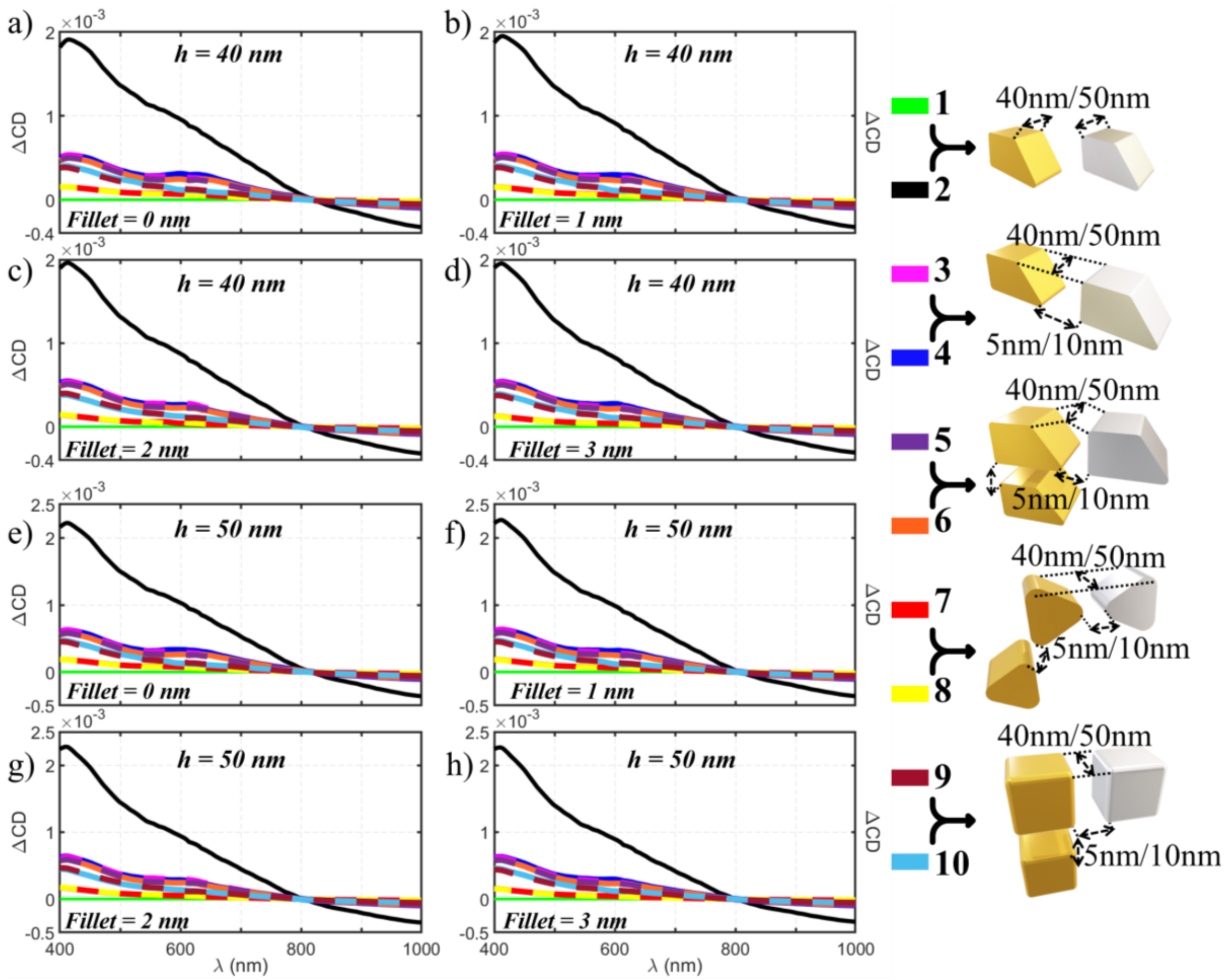

**Fig. S17.** Magnetic tunability of the chiro-optical response, MCD (that is, ΔCD), based on the curvature radius on nanostructure tips; (a,e) without fillets, (b,f) with 1 nm fillets on tips, (c,g) with 2 nm fillets, (d,h) and 3 nm fillets as the curvature radius on tips. The corresponding height of each system is given as $h$ = 40 nm (a-d), or $h$ = 50 nm (e-h), and different inter-nanoparticle gaps for the case of bimetallic nanoantennas as $d$ = 5 nm (dashed lines), and 10 nm (solid lines). Specifically, the studied systems are: (1) monomer of Au, and (2) monomer of Ni trapezoids. Dimer of Au and Ni trapezoids with (3) 5 nm nanoparticle gaps and, (4) 10 nm gaps. Trimer of Au, Au, and Ni trapezoids with (5) 5 nm nanoparticle gaps and, (6) 10 nm gaps. Trimer of Au, Au, and Ni triangles with (7) 5 nm nanoparticle gaps and, (8) 10 nm gaps. Trimer of Au, Au, and Ni cubes with (9) 5 nm nanoparticle gaps and, (10) 10 nm gaps.